\documentclass[lettersize,journal]{IEEEtran}
\usepackage{amsmath,amsfonts}
\usepackage{algorithmic}
\usepackage{algorithm}
\usepackage{array}
 
\usepackage{textcomp}
\usepackage{stfloats}
\usepackage{url}
\usepackage{verbatim}
\usepackage{graphicx}
\usepackage{cite}
\usepackage{booktabs}
\usepackage[commandnameprefix=always]{changes} 
\usepackage{subcaption}
\usepackage{isotope}
\usepackage{comment}
\usepackage[switch]{lineno}

\usepackage{threeparttable}
\usepackage{makecell, multirow}

\begin{document}
\bstctlcite{IEEEexample:BSTcontrol}
%\linenumbers

\title{MERMAID-v1 PET Scanner Prototype: Performance Study and First Zebrafish Scans}

\author{Steven Seeger$^{*}$, \IEEEmembership{Member, IEEE}, Rebecca Kantorek$^{*}$, \IEEEmembership{Graduate Student Member, IEEE}, Hong Phuc Vo, \IEEEmembership{Graduate Student Member, IEEE},  Jorge Roser, \IEEEmembership{Member, IEEE}, Magdalena Ko\l{}odziej, Ezzat Elmoujarkach, \IEEEmembership{Member, IEEE}, Caroline Florack, Christian Schmidt, and Magdalena Rafecas, \IEEEmembership{Senior Member, IEEE}
        % <-this % stops a space
\thanks{$^{*}$S.~Seeger and R.~Kantorek contributed equally to this work.}
\thanks{This work involved animals in its research. Approval of all ethical and experimental procedures and protocols was granted by the \textit{Ministerium für Landwirtschaft, ländliche Räume, Europa und Verbraucherschutz Schleswig Holstein} under NTP-ID: 00030815-4-0, and performed in line with the German animal welfare legislation.}        
\thanks{S.~Seeger was with the Institute of Medical Engineering, Universität zu Lübeck, Lübeck, Germany. He is now with the Isotope Laboratory of the Natural Sciences Section, Universität zu Lübeck, Lübeck, Germany. R.~Kantorek is with the Institute of Medical Engineering and the Isotope Laboratory of the Natural Sciences Section. H.P.~Vo, M.~Ko\l{}odziej, J. Roser and M.~Rafecas are with the Institute of Medical Engineering. E.~Elmoujarkach was with with the Institute of Medical Engineering. He is now with the Radiology Department, Weill Cornell Medical College, New York, United States.
C.~Schmidt is with the Isotope Laboratory of the Natural Sciences Section. C.~ Florack was with the Institute of Medical Engineering. She is now with the Institute of Radiology and Nuclear Medicine, University Hospital Lübeck, Germany.
}

\thanks{The project is partially supported by the German Research Foundation (DFG) under grant agreement no. 496099829 and the DFG Cluster of Excellence PMI under grant agreement no. 390884018.}
}

% The paper headers
\markboth{IEEE TRANSACTIONS ON RADIATION AND PLASMA MEDICAL SCIENCES, VOL. XX, NO. XX, XXX}%
{Shell \MakeLowercase{\textit{seeger et al.}}: MERMAID, a Small Aquatic Animal PET Scanner Prototype}

\maketitle

\begin{abstract}
MERMAID-v1 is a prototype PET scanner designed to support biomedical research involving adult zebrafish and similar species. The current experimental setup has been characterized, and scans of various phantoms, as well as adult zebrafish have been conducted. A dedicated reconstruction software was implemented, 
including models of the degradation effects arising from the lack of depth-of-interaction (DOI) information.
The average energy resolution was 21.6\% (FWHM at 511\,keV), with no significant dead-time effects observed for activities up to 18\,MBq. The absolute sensitivity at the center of the
field of view (FOV) 
ranged from 0.06\% to 0.31\%, depending on the energy window  (from 450-550 to 300-600\,keV), reflecting the limitations of the current two-head configuration. In the central 12\,mm of the transaxial FOV, the averaged spatial resolution is approximately 0.77\,mm (FWHM) transaxially and 0.66\,mm axially, as evaluated using a point source. Image quality was assessed using a downscaled NEMA-inspired IQ phantom and a 3D-printed Derenzo phantom. The reconstructed images of the latter suggest a spatial resolution around 0.7\,mm - 0.8\,mm, despite the lack of DOI information. The first ex- and in-vivo PET scans of adult zebrafish, previously anesthetized fish and placed in a water-filled imaging chamber, were successfully performed, showing detectable tracer uptake in organs such as the brain and eyes despite low initial activity levels. These results confirm MERMAID-v1 capability to obtain useful images from living fish specimens.
While no scatter, attenuation, or efficiency corrections have yet been implemented, this work establishes a working proof-of-concept for dedicated PET imaging of small aquatic vertebrates. Future developments will focus on developing correction techniques, expanding the detector array, and integrating complementary modalities such as computed tomography.
\end{abstract}


\section{Introduction}
\IEEEPARstart{I}{n} recent years, the use of aquatic animals as model organisms in biomedical research has significantly increased, 
especially zebrafish (lat. \textit{Danio rerio}) \cite{Meyers_2018,Liu_2022,Teame2019, zebrafishAdultAnatomyAtlas}.
Optical imaging techniques, e.g. fluorescence microscopy, are widely used to analyze the embryos~\cite{zebrafishEmbryoAnatomyAtlas} and larvae of zebrafish because they are transparent. However, the opacity of adult zebrafish poses some limitations as a result of the limited penetration depth of light. 
In contrast to larger laboratory animals, the spectrum of in-vivo imaging technologies available for adult zebrafish is limited. In particular, the potential use of Positron Emission Tomography (PET) has hardly been tested, even if demand exists.
The few known cases were performed using commercial scanners dedicated to small rodents, for example \cite{Tucker2021, Snay1127, NAZARIO_2026, Lai2026}.  
(The latter cannot be considered an in-vivo study, as the animal was brain dead.) These works underline the increasing interest in zebrafish imaging and the potential of PET. 
A reason behind the limited application of PET to zebrafish  is the small size of the animal (up to 4~cm). 
Additionally, unless the scan is very short, the fish needs to be placed in an aquatic environment and remain fixed as well as anesthetized during the scan.
Addressing these challenges is at the heart of the MERMAID project. MERMAID stands for \textit{Multi-Emission Radioisotopes - Marine Animal Imaging Device} and essentially involves the construction of a dedicated PET prototype for in-vivo imaging of adult zebrafish. Furthermore, the project deals with  
alternative tracer application routes, the construction of a multi-modality imaging chamber \cite{Seeger2022}, the development of 3D-printed specific imaging phantoms to evaluate the prototype \cite{ammm22,Elmoujarkach_2024}, and the integration of computed tomography (CT).
The initial proof-of-concept (PoC) PET prototype was introduced in \cite{Zvolsky2019a}, while \cite{seeger_22_IEEE,seeger_23_IEEE,MERMAID_NSSMIC_25} described some initial results with an upgraded system. 
In this paper, we present a comprehensive study including some of the latest instrumentation developments. Results of the performance evaluation were obtained by adapting the National Electrical Manufacturers Association (NEMA) NU 4-2008 standard \cite{nema}. Given that its prescribed measurements are not fully applicable to the partial-ring, small-bore geometry of MERMAID, the NEMA standard was used as a reference framework rather than as a strict testing protocol. This evaluation was complemented with ex-vivo and in-vivo measurements of adult zebrafish.

\section{Methods}
\subsection{MERMAID PET scanner}
\begin{figure*}
    \centering
    \centering
    \subcaptionbox{} {\includegraphics[width=0.32\linewidth]{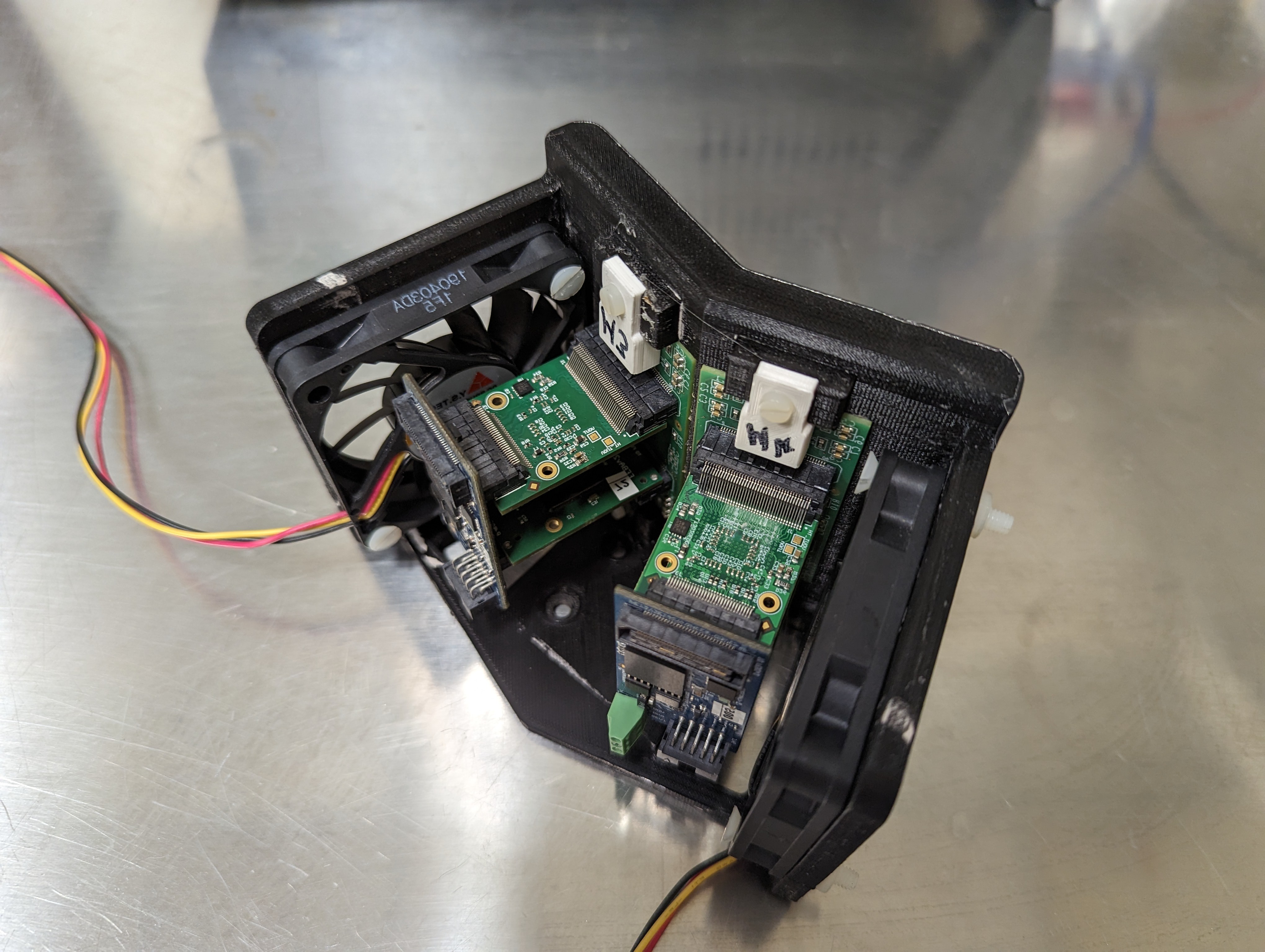}}
    \subcaptionbox{} {\includegraphics[width=0.32\linewidth]{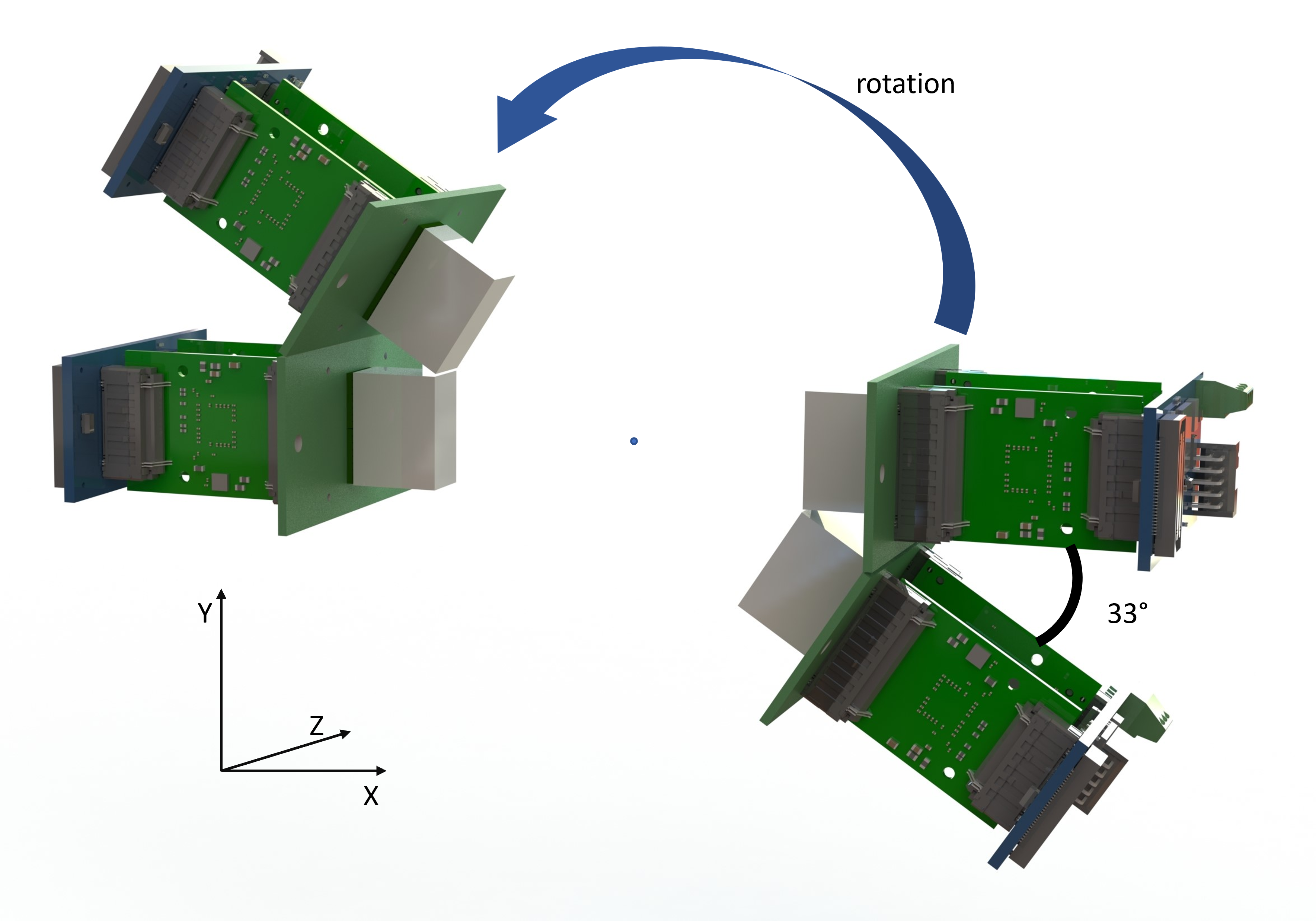}} 
    \subcaptionbox{} {\includegraphics[width=0.32\linewidth]{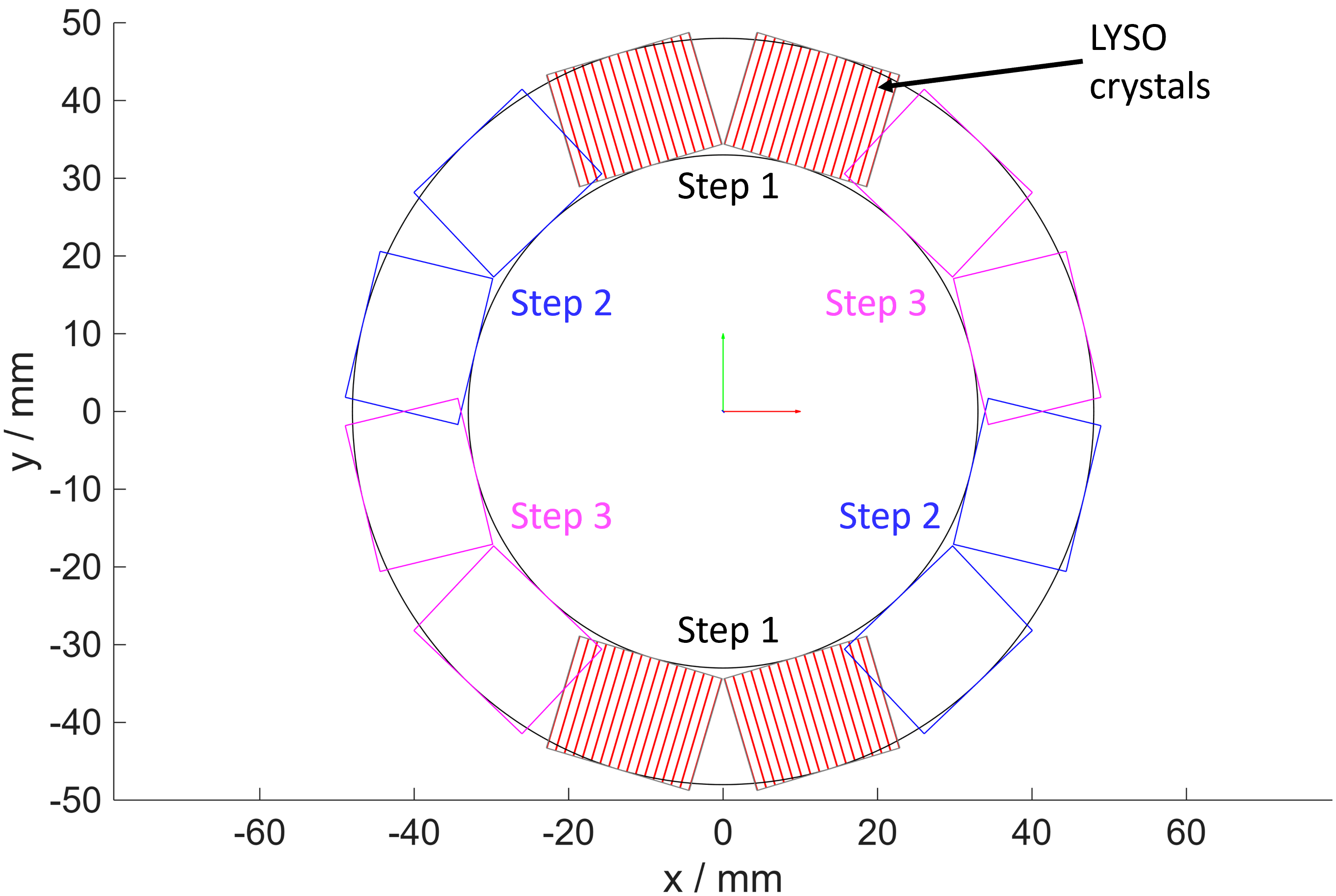}}
    \caption{(a) Detector head, with two detector modules and fans. (b) Rendered image of the detector modules in the current configuration, without holders and fans. (c) 
    Positions of the modules for a 3-step acquisition (default):
    The positions for the three steps are depicted in different colors; for step one, the individual crystal positions in the crystal matrix are shown in red.    
    }
    \vspace{0.5em} 
    \rule{\textwidth}{0.4pt}
    \label{fig:det_all}
\end{figure*}

The current system, MERMAID-v1, consists of four detector modules equally distributed in two detector heads (see Figs. \ref{fig:det_all}(a) and (b)). A detector module is composed of a 128-element matrix of 16$\times$8 pixellated crystals of Lutetium–yttrium oxyorthosilicate doped with Cerium (LYSO:Ce), coupled to two Hamamatsu Silicon Photomultipliers (SiPM) (S13615-1050N-08) which are placed next to each other on a common circuit board. 
The individual crystals measure 1.12$\times$1.12$\times$15.00\,mm$^3$ with a pitch of 1.2\,mm, and are coupled one-to-one to the SiPM units.
This configuration yields a total of 128 detector channels with 396 pixels per channel (with a pitch of 50\, \textmu m).  The crystals are separated from each other by means of a ESR reflector film (3M,  0.07\,mm), and were used as supplied by the manufacturer. 
The use of ESR was motivated by its high reflectivity and efficient light collection, as previously exploited in the EndoTOFPET-US detector concept \cite{FRISCH2013577}. 
The optical coupling between crystals and SiPMs was done using a 25\,\textmu m layer of optical glue (DOWSIL 3145 RTV Mil-A-46146). For the SiPM readout, we use two PETsys TOFPET 2C ASIC \cite{tofpet2} per detector module, as well as the corresponding readout electronics. The two ASIC are connected to the PCB housing the SiPMs at a 90\textdegree\ angle. This design allows for simple way of cooling the ASICs using fans. 

Within one detector head, the two modules are placed at 33\textdegree\ to each other, so each module in one head is parallel to a corresponding module in the other head  for the distance specified above, see Fig.~\ref{fig:det_all}(b).
To scan an object,
the detector heads rotate over 180\textdegree\  and acquire data in a step-and-shoot mode.
The distance between opposite detectors as well as the number of rotation steps is adjustable, although the design of the heads has been optimized for a module-to-module distance of 66\,mm. This distance  allows for the placement of the small-fish imaging chamber (see section \ref{subsec:chamber}). 

As part of the system characterization, different rotation step numbers and angle values were tested.
As a compromise between acquisition time and image quality, 
three steps (60\textdegree\ per step) were chosen as default parameter.  The corresponding positions are illustrated in Fig. \ref{fig:det_all}(c).

To scan objects larger than the axial extend of the modules (9.6\,mm),
a motorized linear stage with position encoding is used (0.01\,mm precision). While the number of axial steps per scan is adjustable, the default imaging protocol employs 3\,mm displacements between positions. Note that both the rotation and the axial steps lead to some overlap of the modules' positions. 
In order to compensate for radioactive decay, the acquisition time at each rotation and axial step is
progressively extended to keep a constant mean number of disintegrations per step. 

\begin{figure}
    \centering
    \includegraphics[width=0.2\textwidth]{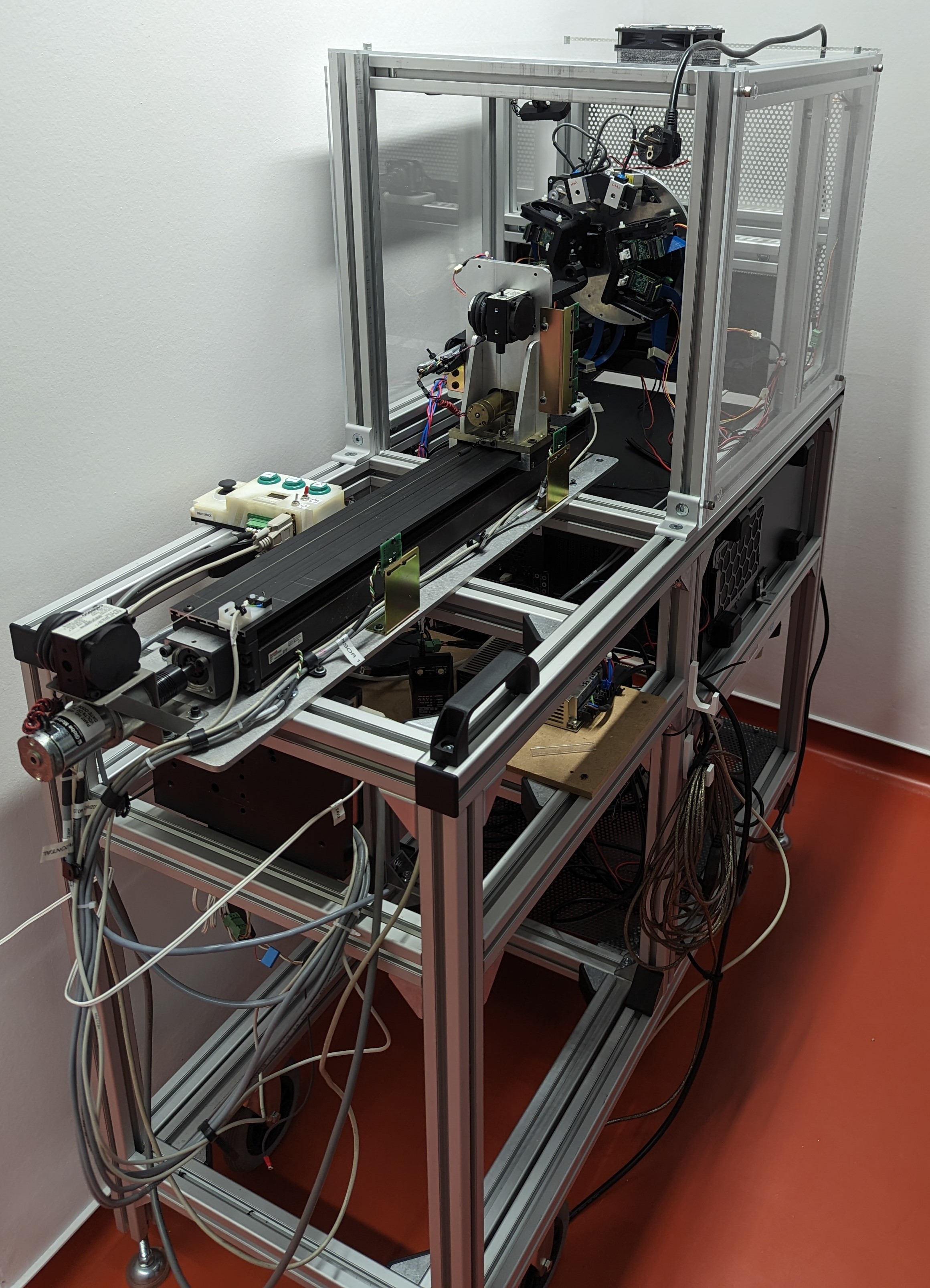}
    \caption{Current status of the MERMAID-v1 prototype.}
    \label{fig:prototype_laboratory}
\end{figure}

The rotating stage with the detectors, as well as a linear stage for probe positioning, was mounted in an aluminum rack together with  all readout electronics, the data-acquisition computer, and power supplies. The entire frame is placed on wheels for flexible use and transportation, see Fig.~\ref{fig:prototype_laboratory}.

\subsection{ASIC calibration and threshold optimization}
The SiPMs are operated at a breakdown voltage (V$_{br}$) of 53\,V and an overvoltage (V$_{ov}$) of 3\,V. The impact of different V$_{ov}$ on the energy spectrum and photopeak position was evaluated in previous studies \cite{seeger_22_IEEE}.
The ASICs were first calibrated with the standard PETsys calibration routine~\cite{softwareUserGuidePETsys}. 
 
The goal of threshold optimization was to obtain a distinct spectrum for the measured sources in which the corresponding photopeaks were clearly recognizable. Given the low count rates expected, and the small bore size, no particular emphasis was placed on optimizing the temporal resolution. The event trigger logic of the TOFPET2C ASICs has three relevant threshold voltages (V$_{T1}$, V$_{T2}$ and V$_{E}$) that determine the acceptance of a signal~\cite{Bugalho2019}. Each one is controlled by a global multiplier and a channel-specific value ($vth\_t1$, $vth\_t2$ and $vth\_e$, respectively). The global multiplier is set before the initial calibration described above. The local thresholds need to be determined according to the gain and noise level of each detector channel. Since the values of $vth\_t1$ and $vth\_t2$ mainly contribute to the timing resolution, the main goals of the threshold optimization were (1) to find a $vth\_e$ value which led to a reasonable energy spectrum, i.e., with defined fraction of photopeak events compared to background for the measured source  (default: 40\%), and (2) to suppress the fraction of noise events not related to a photon detection.
To select the threshold values, a semi-automatic algorithm was implemented. This algorithm evaluates the measured energy spectra and determines the ratio between the single events in the photopeak compared to all other events. Photopeak events are defined as those that lie within the doubled full-width at half maximum (FWHM) of the photopeak Gaussian fit. The threshold values are then adjusted per channel until a ratio of 20-50\% is achieved.  

\subsection{Energy calibration}
Measurements of $^{133}$Ba, $^{22}$Na and $^{137}$Cs point sources placed at the center of the scanner were taken for 1800\,s each. Activities at time of calibration were 624, 370 and 942\,kBq for $^{133}$Ba, $^{22}$Na and $^{137}$Cs, respectively. The relevant energy peaks of 356\,keV, 511\,keV and 662\,keV were fitted with Gaussian functions, and the position of the photopeak maxima after the fit were used as a calibration points; the FWHM of each photopeak fit served as an estimate of the uncertainty affecting the corresponding point towards fitting the calibration function. For the latter, we implemented 
\begin{equation}
	f(x)=a(1-e^{-\frac{bx}{a}}),
	\label{eq:ekali}
\end{equation} 
where $x$ is measured photopeak position in ADC units, while $a$ and $b$ are free parameters. This function results from adapting the formula presented in \cite{Roncali2011} to describe the behavior of a SiPM for different photon energies.      
Next, the energy resolution was evaluated as FWHM of the 511\,keV photopeak from measurements of $^{18}$F-FDG, $^{22}$Na, and $^{89}$Zr sources and phantoms in order to study possible dependencies of energy resolution on the radioisotopes with additional gamma emissions. In the case of  $^{22}$Na, the  source described above was employed. For $^{18}$F-FDG and $^{89}$Zr, a 3D printed downscaled NEMA NU4 Phantom (see section \ref{meth_spat_res}) was used, with activities of 13.1 MBq 
($^{18}$F-FDG) and 4.2 MBq ($^{89}$Zr). 

\subsection{Data processing and image reconstruction}
\label{subsec:reco}
A custom algorithm was used to sort coincidences from the list of energy-calibrated single events,
on the basis of a pre-set time and energy windows. Only  coincidences consisting of two single events in opposed heads were considered, i.e., inter-crystal scatter events were excluded. Without optimization, the coincidence timing resolution was 920\,ps, so a  
time window 
of 2\,ns was applied. Unless otherwise specified,  the energy window spanned from 
 450 to 550\,keV. This window was chosen to reduce the amount of scatter events.
The resulting coincidence list includes
times, energies, and the channel numbers in the modules. 

For image reconstruction, we have implemented the
list-mode maximum-likelihood expectation-maximization (LM-MLEM) algorithm. 
The required system matrix and sensitivity image  were calculated using a multiray approach similar to \cite{Gillam_2013, vo2025dedicated}.  
For a given coincidence, several rays  connecting the involved detectors are randomly traced, with their endpoints distributed uniformly within the crystal volume. Attenuation in the crystal is  taken into account to model crystal penetration. This approach enables modeling of the Volume-of-Response (VOR)  to compensate for parallax errors.
Neither attenuation nor scatter corrections have yet been implemented. 
Efficiency corrections were not applied.

The reconstruction algorithm jointly processes the data taken at all axial steps to generate a single 3D image. To this aim, the system matrix and sensitivity image required for reconstruction were accordingly adapted.
The voxel and image-grid size were selected depending on the specific imaged object. For the resolution study described below, an image difference below 1\% was used as stopping criterion.
Otherwise, the number of iterations was chosen based on visual inspection of the results. The reconstruction parameters are summarized in table~\ref{tab:reco_params}. Whereas quantitative image analysis was performed using the raw images, a median filter with a kernel of 3$\times$3$\times$3 voxels was applied for visualization and display (except for the Derenzo phantom analysis, see below). 

\subsection{Measurements and phantoms}
In addition to point sources, several phantoms were also used in this work for the following purposes.  

\subsubsection{Count-rate performance}
A 3D-printed hollow cylinder was filled with $^{18}$F-FDG. The phantom, with an active diameter of 7\,mm and a length of 20\,mm, was placed in the center of the scanner and  measured over 315\,min, providing 63 single measurements of 3\,min each. The initial activity was 18.42\,MBq. Such homogeneously-filled phantom, which deviates from the line source recommended by the NEMA protocol, was purposefully chosen to mimic more closely the actual imaging scenario of adult zebrafish. The number of events as well as the number of coincidences per measurement was plotted against the corresponding activity.

\subsubsection{System sensitivity}  
\label{meth_sensitivity}
A  $^{22}$Na point-like source   with a diameter if 0.25\,mm, an activity of 0.72\,MBq and placed within an acrylic cube of 1\,cm$^{3}$ was moved along the z-axis of the system in 0.5-mm steps.
This step size deviates from the NEMA protocol, which specifies a step size equal to the slice thickness of the reconstructed images (0.25\,mm or smaller in  our case); this deviation was motivated by the limited precision of  the
available linear stage.
Concerning the acquisition time, the NEMA protocol recommends a scan length that provides a minimum of 10\,000 true coincidences. Since the number of random and scatter events that MERMAID-v1 detects could not be estimated at that stage, 
we measured for 300\,s per position to acquire a sufficiently large number of prompt events. This resulted in 112\,622 coincidences for the default energy window and the source at the center of the FOV. 
The absolute sensitivity $S_{a}$ was determined according to Eq.~\ref{eq:sens}~\cite{nema} 
\begin{equation}\label{eq:sens}
   S_{a}(\%)=\frac{\frac{R_{C}-R_{B}}{A}}{0.9060} \cdot100,
   %\nonumber
\end{equation}
where $R_{C}$ and $R_{B}$ are the rates of coincidences per second measured with and without source, respectively, while $A$ is the activity of the source in Bq; the value in the denominator corresponds to the branching ratio of $^{22}$Na for positron emission.
To estimate the effect of the energy window  on the sensitivity,
three  windows widths were considered:
from 450\,keV to 550\,keV (default), as well as two larger windows, namely from 400\,keV to 600\,keV, and from 300\,keV to 600\,keV.

\subsubsection{Spatial resolution and image quality}
\label{meth_spat_res}
We have adapted the approach from the NEMA NU 4-2008 standard (conceived for mouse and rat imaging) 
to MERMAID (small-fish imaging with very small FOV). 
The $^{22}$Na point source (same as in section \ref{meth_sensitivity}) was placed at different positions along the x- and y-axes of the device (see coordinate reference system in Fig.~\ref{fig:det_all}). Starting at the center, the source was moved in 1-mm steps over 10\,mm in x- and y-direction respectively. 
The acquisition time was 300\,s per step. 
For each source position and each spatial dimension, the
FWHM 
was calculated from the reconstructed images at iteration 20,
after summing over all slices in the two remaining spatial dimensions, using a linear interpolation between adjacent voxel values. 

\begin{figure}[htb]
    \centering
    \includegraphics[width=0.4\textwidth]{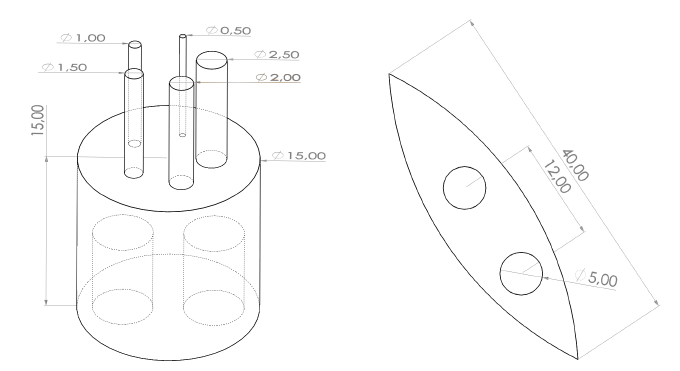}
    \caption{Schematic view of the downscaled IQ phantom (left) and central section of the fish-like phantom (right), with dimensions in millimeter}
    \label{fig:phants}
\end{figure}

An Image Quality (IQ) Phantom  was 3D-printed using $^{18}$F-FDG mixed with resin, following the procedure described in \cite{ieee22_3d}. The printer provides a precision of around 0.035\,mm in the xy-plane of the printer.
This phantom was inspired by NEMA NU4 IQ phantom design \cite{nema}, and downscaled by  50\% (see Fig.~\ref{fig:phants}). Its  diameter was 15\,mm, with a total length of 30\,mm. The rod region consisted of five rods with diameters ranging from 0.5 to 2.5\,mm. The uniform region was 7.5\,mm in length.
The bottom part included two air-filled cylinders of 7.5\,mm length and a diameter of 4\,mm.
The phantom was scanned at 9 axial positions; 
the total measurement time was 3\,201\,s, with an initial activity concentration of  1.74\,MBq/ml.
This deviation from the NEMA standard was made to accommodate the low sensitivity of the current partial system, as well as the current lack of random and scatter corrections.

The image quality was evaluated by calculating the recovery coefficient (RC) for each rod, the uniformity $U$, and the spillover ratio (SOR) according to the NEMA standard, with
\begin{equation}
	RC=\frac{\mu_{rod}}{\mu_{hom}}, ~~~ U=\frac{\sigma_{hom}}{\mu_{hom}}	~~~ \text{and}~~~ SOR=\frac{\mu_{ROIs}}{\mu_{hom}};\nonumber
\end{equation} 
${\mu_{rod}}$ and ${\mu_{hom}}$ are the maximum pixel value in a 5\,mm long region-of-interest (ROI) with a diameter double the rod diameter,
and the mean of the homogeneous phantom part, respectively. 
The latter measured 5 mm in height, with a diameter of 11.25\,mm (75\% of the nominal diameter). This ROI was also used to calculate the uniformity, with  $\sigma_{hom}$ being the standard deviation of the intensity values within that ROI.
For the SOR, ${\mu_{ROIs}}$ corresponded to the mean value in two ROIs, each placed in one of the two air-filled cylinders, with a diameter of 2\,mm (50\% of the nominal diameter) and a length of 3\,mm.

\begin{figure}[htb]
	\centering
    \subcaptionbox{} {\includegraphics[width=0.49\linewidth]{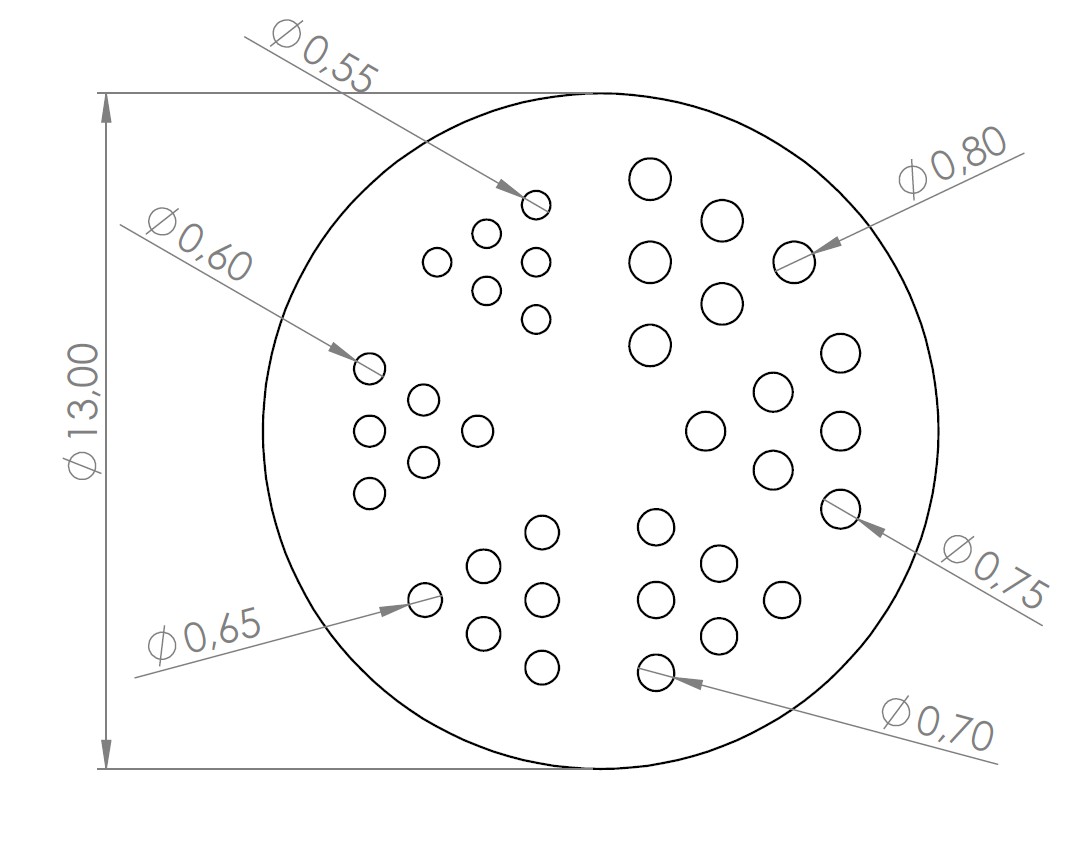}} 
	\subcaptionbox{} {\includegraphics[width=0.49\linewidth]{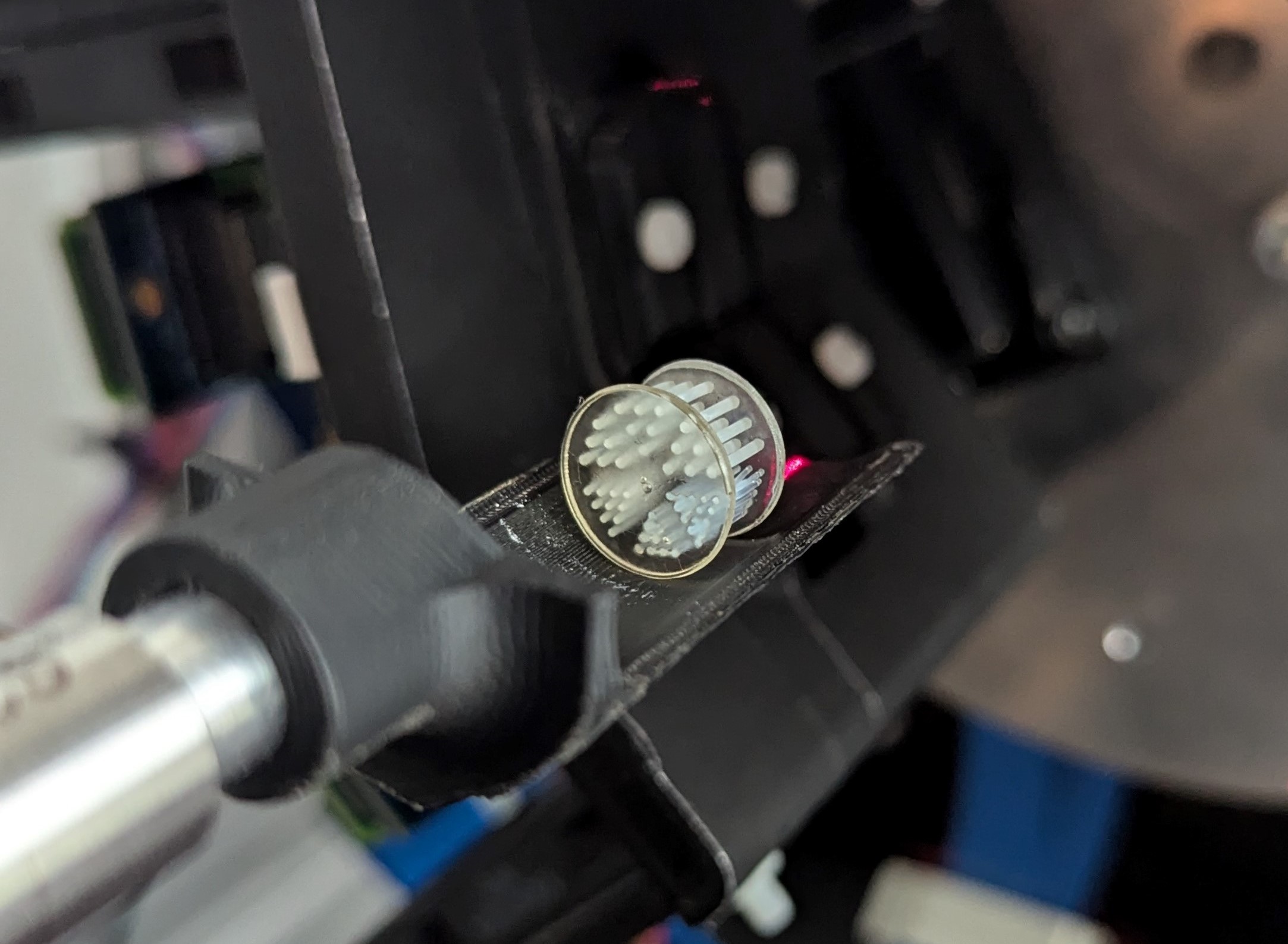}}
	\caption[]{(a) Design of the Derenzo phantom, size in mm. (b) Derenzo phantom placed on sample holder prior to a scan. Behind the phantom, one of the detector heads is visible.}
	\label{fig:derenzo}
\end{figure}

Furthermore, a Derenzo phantom (Fig. \ref{fig:derenzo}) was 3D-printed using $^{18}$F-FDG mixed with resin. The phantom, inspired by \cite{Kang978}, has a  diameter of 13\,mm and  rod of various sizes distributed in six sectors, with rod diameters  0.55, 0.60, 0.65, 0.70, 0.75 and 0.80\,mm, respectively. The activity concentration at the start of the measurement was 8.09\,MBq/ml. Note that the active rods of the phantom were surrounded by air, i.e., without non-radioactive walls. After reconstruction, the transverse slices were projected along 6\,mm in the axial direction to generate a summed image, to decouple the effect of statistical noise in the spatial resolution (similarly to  \cite{Kang978}). From this projection image, the resolvability of each phantom sector was evaluated according to \cite{Hallen2020,kang2026sub}: Line profiles were defined over three different orientations, with  
valley-to-peak ratios (VPR)
computed for each profile; the resolvability was calculated as:
\begin{equation}
    \mathrm{Resolvability}(\%) =\frac{N_{\mathrm{Rayleigh}}}{N_{\mathrm{total}}}
\times 100,
\end{equation}
where $N_{\mathrm{Rayleigh}}$ is the number of evaluated peak–valley measurements satisfying the Rayleigh criterion, namely $
\mathrm{VPR}<0.735$, whereas $N_{\mathrm{total}}$ is the total number of evaluated measurements in a rod sector.
This analysis was performed on the image of the Derenzo phantom obtained at iteration 100 and post-processed with a median-filter to control noise amplification.  

The image quality metrics from the IQ phantom were evaluated as a function of the iteration number. 

\subsubsection{Fish-like phantom}
As a surrogate of a real zebrafish, 
a fish-like phantom was 3D-printed, similarly to \cite{Elmoujarkach_2024}. 
This phantom was designed to optimize data acquisition protocols for zebrafish, following  the principle of \textit{Replacement, Reduction, and Refinement} (3R)  for animal experiments. Furthermore, it was used to assess and validate reconstruction and acquisition protocols for elongated objects, with compensation for radioactive decay during  rotation and z-axis movement during the scan.
The phantom consists of two 3D-printed halves that are joined together to form a fusiform structure, the ``fish body'' (see Fig.~\ref{fig:phants}, right). Each half is solid except for two spherical cavities (diameter: 5\,mm). 
The activity concentration in the fish's body was  2.57\,MBq/ml. 
The phantom was scanned in 14 axial positions 
(total acquisition time 65\,min).

\subsection{Imaging chamber}
\label{subsec:chamber}
The imaging chamber is  essential to  ensure that the fish remains anesthetized and immobilized in an aqueous environment during the scan. As animal welfare is the top priority, the chamber was designed to prevent  
injuries caused by the immobilization holder and to minimize additional stress during the measurement. 
The imaging chamber consists of a transparent cylinder made of polymethyl methacrylate (PMMA) with an outer diameter of 26\,mm (wall thickness of 2\,mm) and a length of 144\,mm. This new design improves upon the former version published in \cite{Seeger2022} by being further downscaled, thereby reducing the amount of attenuating material within the FOV.
The transparent cylinder allows visual inspection of the overall condition of the animal as well as certain vital signs, such as gill movement. 
Fresh water and anesthetics were pumped through the chamber with a flow of 0.16\,ml/min. To quickly place the fish and the immobilization holder in the cylinder, the upper part of the cylinder is removable but sealed when in place. The immobilization holder was a flexible 3D printed structure or a sponge.

\subsection{Ex-vivo \& in-vivo zebrafish scans}
Five adult zebrafish
were scanned in-vivo\footnote{Zebrafish husbandry and experiments were performed following German animal welfare legislation (NTP-ID: 00030815-4-0) under the supervision of the local representative of the animal welfare agency.}. Each fish was anesthetized using tricain-methansulfonat (MS‐222); 5\,$\mu$l of $^{18}$F-FDG were injected intraperitoneally (IP) in the lower belly, followed by 30-min awake incubation in freshwater. In previous tests involving tracer injection with additional five individuals, values of up to 2.27\,MBq were achieved in a single animal; however, these were only allowed to be scanned ex-vivo due to animal testing regulations. 
For the in-vivo experiments, the procedure was the same  until the end of the incubation period. 
The corresponding specimens  showed a uptake between 0.198 and 0.620\,MBq before the scan, depending on the specimen. 
After incubation, the fish were anesthetized again and transferred to the imaging chamber. 
The scans comprised 10 axial positions, with 180\,s for the first axial step, 
thus leading to acquisition times of approximately 45 minutes.  The temperature of the water chamber was not tracked; however, the laboratory conditions during the acquisition were stable with air temperature of 21°C.

\begin{table}[]
	\centering
	\caption[]{Reconstruction parameters for all reconstructed objects.}
	\begin{tabular}{lccc}
		\toprule
                        Object     &  Iterations         & Voxel Size [mm] & Number of voxels \\ \midrule
$^{22}$Na point source & 20 & 0.25 & 150 $\times$ 150 $\times$ 60      \\
IQ phantom & 20 & 0.05 & 550 $\times$ 550 $\times$ 450 \\
Derenzo phantom & 100 & 0.05  & 550 $\times$ 550 $\times$ 240 \\ Fish-like phantom & 20 & 0.25 & 150 $\times$ 150 $\times$ 180 \\ Zebrafish (ex-vivo) & 20 & 0.25 & 150 $\times$ 150 $\times$ 180\\ Zebrafish (in-vivo) & 20 & 0.25 & 150 $\times$ 150 $\times$ 180 
    \\ \bottomrule
	\end{tabular}
	\label{tab:reco_params}
\end{table}

\section{Results}
\subsection{Count-rate performance}
\begin{figure}[htb]
    \centering
    \subcaptionbox{} {\includegraphics[width=0.24\textwidth]{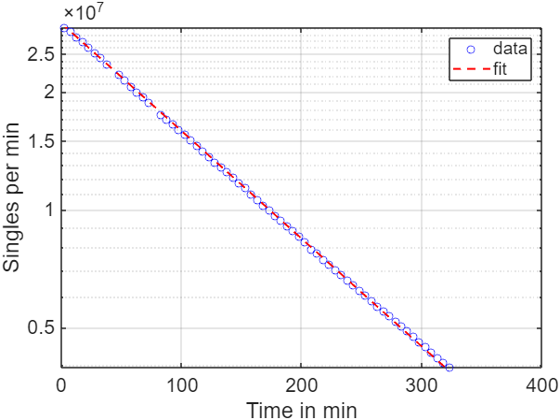}}
    \subcaptionbox{} {\includegraphics[width=0.24\textwidth]{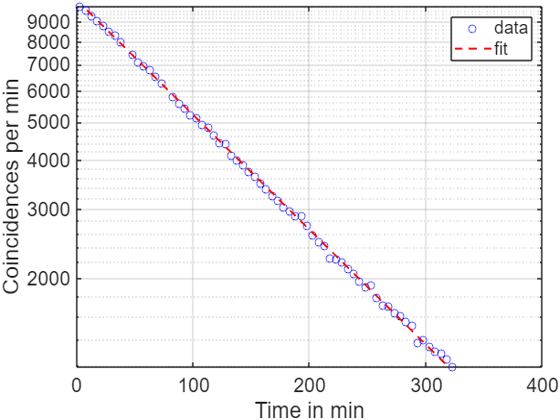}}
    \subcaptionbox{} {\includegraphics[width=0.24\textwidth]{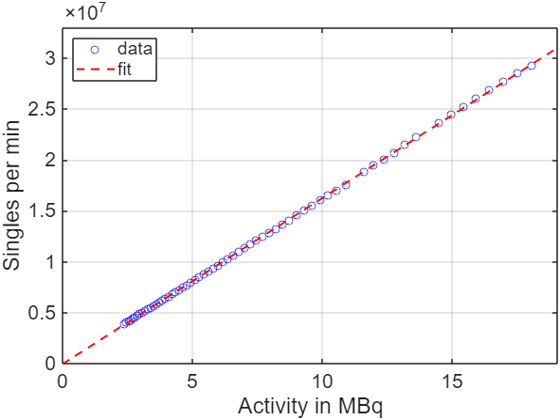}}
    \subcaptionbox{} {\includegraphics[width=0.24\textwidth]{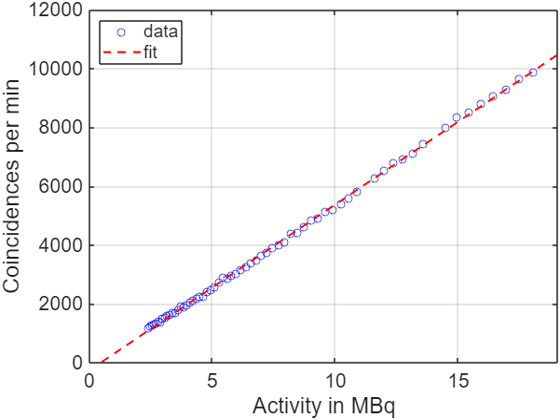}}
    \caption{Count-rate performance of the MERMAID detectors for single events (a,c) and coincidences (b,d) for an activity range from 2 to 18 MBq.}
    \label{fig:coutrate}
\end{figure}

Fig.~\ref{fig:coutrate}  shows the count-rates for singles and coincidences as a function of the activity and time. The rates are expected to exponentially drop as time progresses, which is expressed by the logarithmic y-axis in a straight line. For singles, such exponential fit yielded R$^2$=0.99 (see Fig.~\ref{fig:coutrate} (a)); a half-life of 109.25\,min  was extracted from the fit, which corresponds to a deviation of 0.047\% compared to the literature (109.77\,min). For coincidences, while the fit yielded R$^2$=0.99 (see Fig.~\ref{fig:coutrate} (b)), the extracted half-life was 102.90\,min, a deviation of approximately 6\% compared to literature; additionally, a small offset at zero activity was observed. 

\subsection{Energy calibration \& resolution}
\begin{figure*}
    \centering
    \centering
    \subcaptionbox{\isotope[18]{F}} {\includegraphics[width=0.32\linewidth]{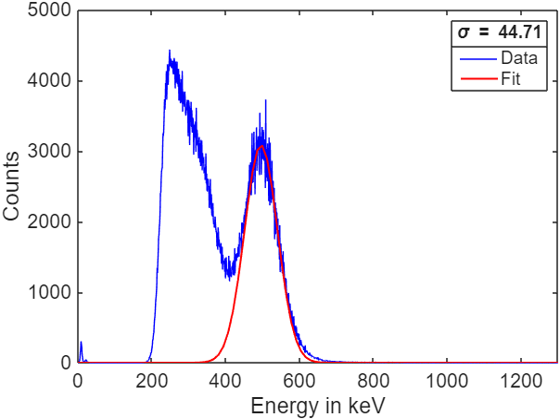}} 
    \subcaptionbox{\isotope[22]{Na}} {\includegraphics[width=0.32\linewidth]{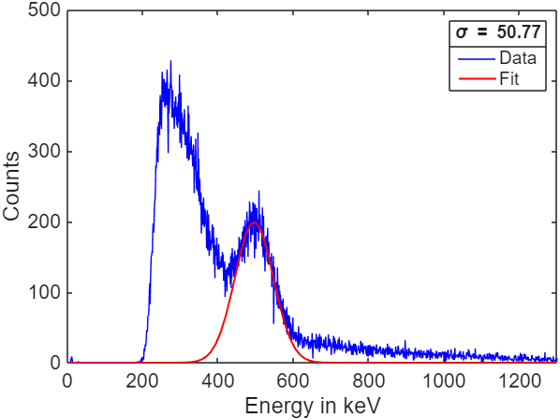}}
    \subcaptionbox{\isotope[89]{Zr}} {\includegraphics[width=0.32\linewidth]{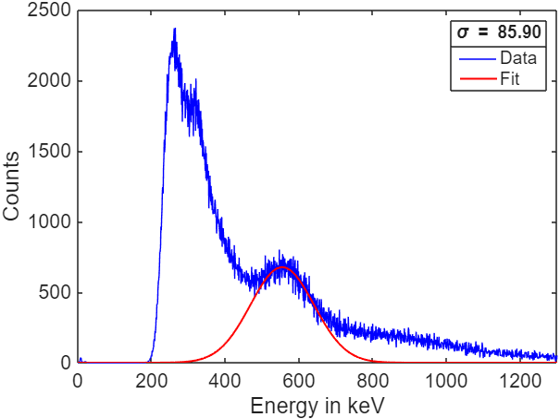}}
    \caption{Energy spectra for \isotope[18]{F},\isotope[22]{Na} and \isotope[89]{Zr} for a representative detector channel. Standard deviation $\sigma$ is expressed in keV. The small peaks near zero energy are pedestal events related to readout electronics, not a physical process, and they have no influence on the analysis.}
    \vspace{0.2em} 
    \rule{\textwidth}{0.4pt} 
    \label{fig:spectra}
\end{figure*}
\begin{figure}
    \centering
    \includegraphics[width=0.9\linewidth]{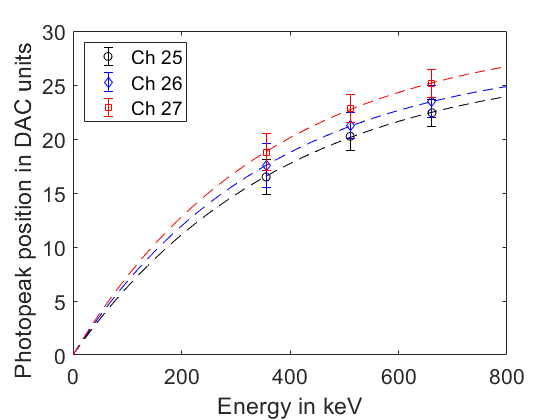}
    \caption{Energy calibration (dashed lines) based on measurements of \isotope[133]{Ba}, \isotope[22]{Na} and \isotope[137]{Cs} for three different channels.}
    \label{fig:caliCurve}
\end{figure}

Energy calibration curves for three representative channels are displayed in Fig. \ref{fig:caliCurve}, while Fig. \ref{fig:spectra} shows the energy spectra after energy calibration for \isotope[18]{F}, \isotope[22]{Na} and \isotope[89]{Zr} for an exemplary channel. The mean and standard deviation of the energy resolution are presented in Table \ref{tab:Eres} for the tested nuclides.
\begin{table}
    \centering
    \caption{Mean energy resolution $\overline{E_{res}}$ of the system for 511\,keV and its standard deviation for the three nuclides. The mean photopeak position $\overline{Pos}$ is shown as well.}
	\label{tab:Eres}
    \begin{tabular}{lcccc}
		\toprule
		Nuclide & \multicolumn{1}{l}{$E_{res}$ {[}\%{]}} & \multicolumn{1}{l}{$\sigma_{r}$ [\%]} & \multicolumn{1}{l}{$\overline{Pos}$ {[}keV{]}} \\ \midrule
		\isotope[18]{F}   & 21.6  & 3.5  & 497.2\\ 
		\isotope[22]{Na}   & 24.9  & 4.0  & 499.7\\  
		\isotope[89]{Zr}  & 44.4  & 16.8  & 560.05\\ \bottomrule  
	\end{tabular}
\end{table}
The curves reflect the strong saturation of the SiPMs at high photon energies.
Compared to \isotope[18]{F}, the downscattering of the 1275-keV gamma emissions broadens the photopeak in the \isotope[22]{Na} spectrum, even if a second peak at this energy cannot be recognized due to the aforementioned saturation. Accordingly, the energy resolution worsens.
These effects were stronger for \isotope[89]{Zr} and its 909\, keV peak because of the high number of 909-keV photons emitted per decay (approx. 99\%), while the branching ratio of positron emission, responsible for the 511-keV peak, is 22.8\%. 

\subsection{Sensitivity} 
\begin{figure}
    \centering
    \includegraphics[width=0.45\textwidth]{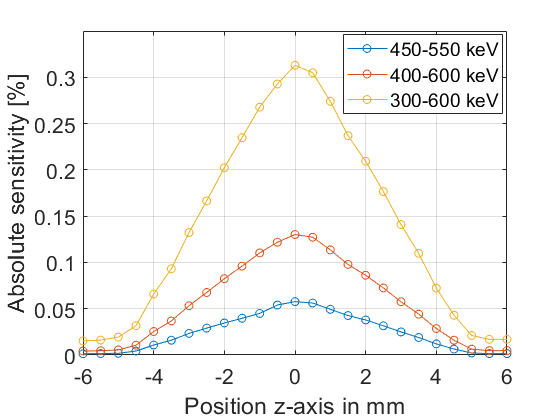}
    \caption{Absolute sensitivity for three different energy windows, measured with a $^{22}$Na in 0.5\,mm steps along the z-axis.}
    \label{fig:resSens}
\end{figure}

The sensitivity profiles for a single axial step are shown in Fig. \ref{fig:resSens}. The maximum sensitivity with the widest energy window is 0.31\%, whereas  it decreases to 0.06\% for the narrowest (default) energy window. 
At the edges, the sensitivity has already dropped to 18.6\% of the maximum.
Nevertheless, only about 8\,mm in the axial dimension are used for reconstruction.
In a scan, the drop in sensitivity is compensated  by using axial steps with overlapping.
\begin{figure*}[htb]
    \centering
    \centering
    \subcaptionbox{RC} {\includegraphics[width=0.32\linewidth]{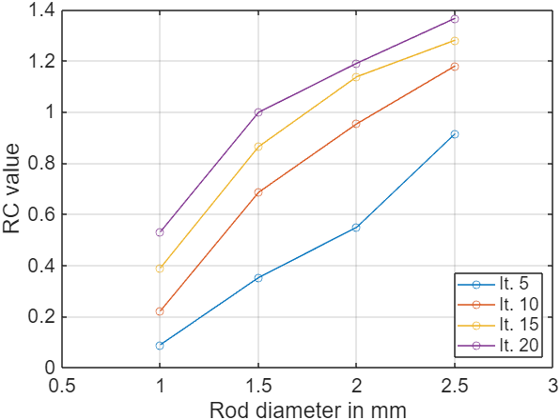}} 
    \subcaptionbox{Uniformity} {\includegraphics[width=0.32\linewidth]{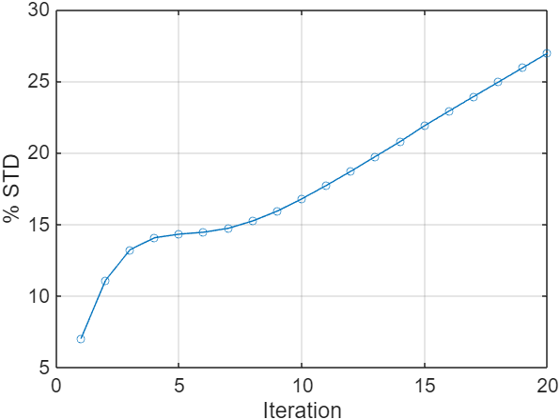}}
    \subcaptionbox{SOR} {\includegraphics[width=0.32\linewidth]{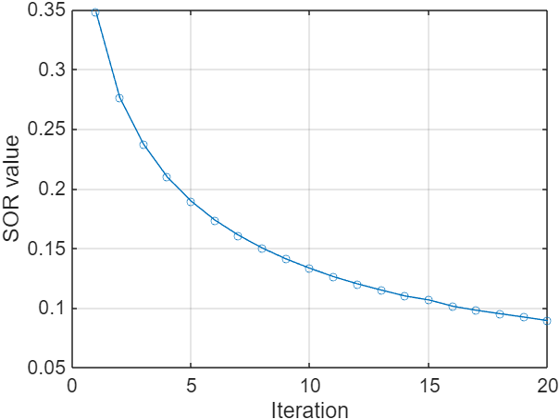}}
    \caption{(a) Recovery coefficient values for iterations 5, 10, 15 and 20; (b) uniformity and (c) spill over ratio calculated from the reconstructed image of the downsized 3D-printed NEMA phantom for several iterations between 5 to 20.}
    \vspace{0.2em} 
    \rule{\textwidth}{0.4pt}
    \label{fig:NEMA_results}
\end{figure*}

\begin{table}[]
	\centering
	\caption[]{Mean (in mm) and relative standard deviation of the FWHM and FWTM for source shifts in the relevant range from 0 to 6\,mm.}
	\begin{tabular}{lccccc}
		\toprule
                                 &             & FWHM & \% Std. & FWTM  & \% Std. \\ 
                                 \midrule
                {X-shift} & Mean x-axis & 0.77          & 2.8   & 1.63 & 4.8   \\
                          & Mean y-axis & 0.82          & 3.6   & 1.73 & 3.9   \\
                          & Mean z-axis & 0.71          & 7.2   & 1.43 & 2.7   \\ \midrule
                {Y-shift} & Mean x-axis & 0.74          & 4.1   & 1.47 & 3.1   \\
                          & Mean y-axis & 0.76          & 6.5   & 1.51 & 1.9  \\
                          & Mean z-axis & 0.61          & 16  & 1.20 & 1.2   \\ 
                                  \bottomrule
	\end{tabular}
	\label{tab:NEMA_FWHM_FWTM}
\end{table}

\subsection{Spatial resolution and image quality}

\begin{figure}[htb]
    \centering
    \subcaptionbox{x-axis shift}{\includegraphics[width=0.35\textwidth]{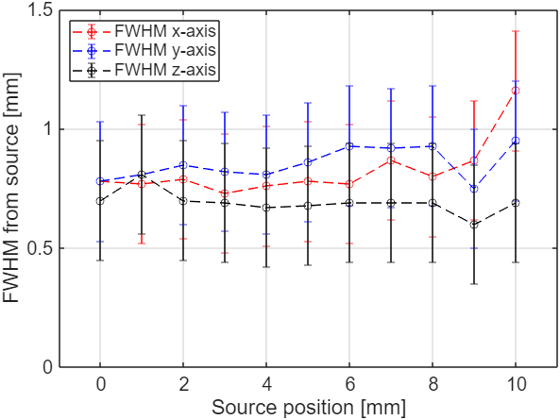}}\label{fig:x}
    \subcaptionbox{y-axis shift}{\includegraphics[width=0.35\textwidth]{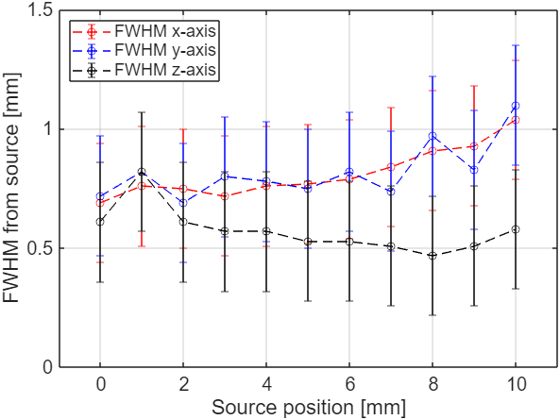}}\label{fig:y}
    \caption{Spatial resolution (FWHM) from measurements of the \isotope[22]{Na} point source, with error bars representing the maximal error bonds assuming the voxel size as the most relevant source of uncertainty. Reconstruction at iteration 20}
    \label{fig:resResolutionFWHMoverview}
\end{figure}

Fig.~\ref{fig:resResolutionFWHMoverview} shows the results for the \isotope[22]{Na} point source measured at 11 different positions along the x- and y-axes of the system. The overall average spatial resolution in the central 12 mm of the transaxial FOV was of 0.77\,mm along the x- and y-axis and 0.66\,mm along the z-axis (see Table~\ref{tab:NEMA_FWHM_FWTM}).

\begin{figure}[t]
    \centering
    \centering
    \subcaptionbox{} {\includegraphics[width=0.49\linewidth]{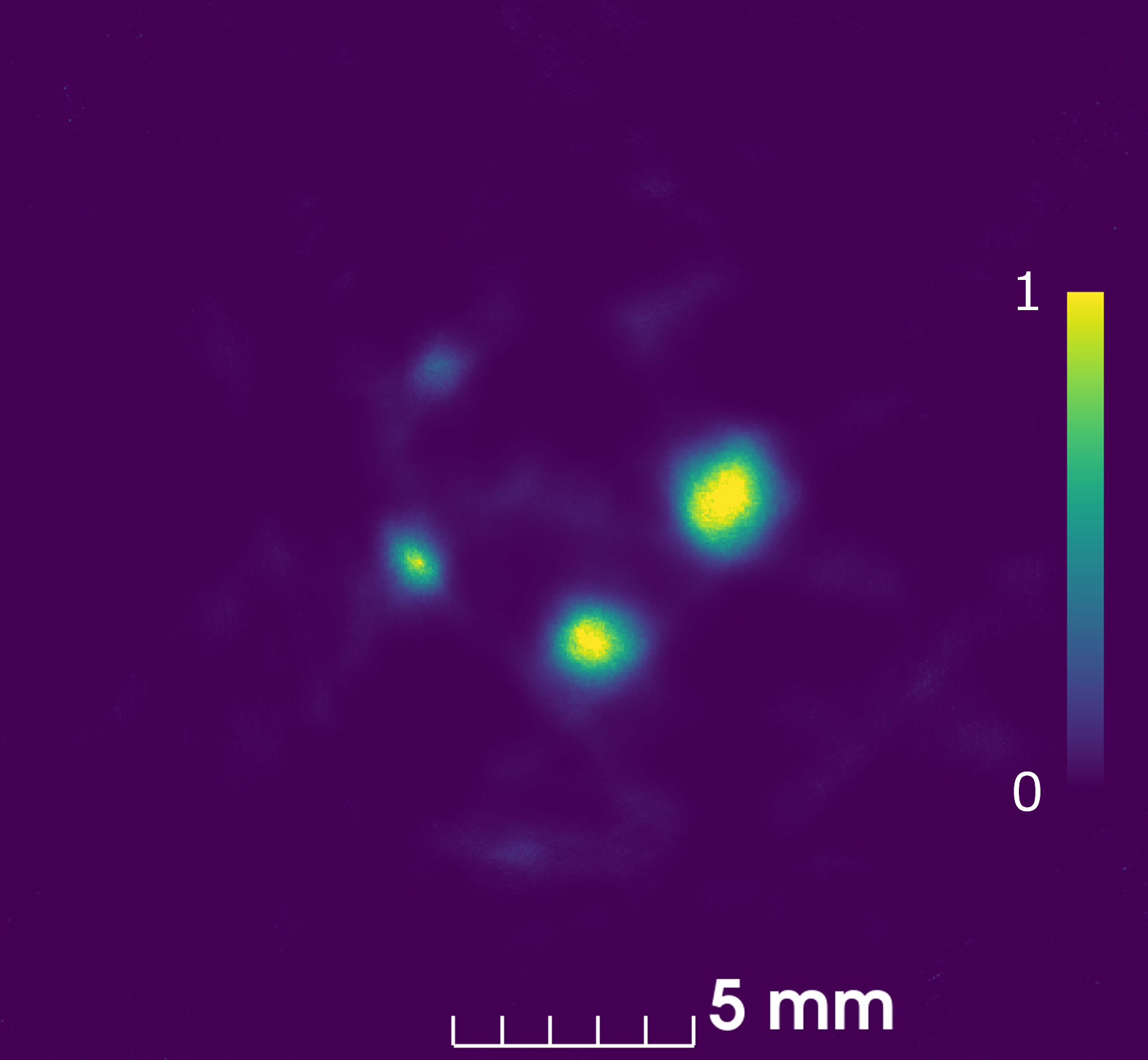}} 
    \subcaptionbox{} {\includegraphics[width=0.49\linewidth]{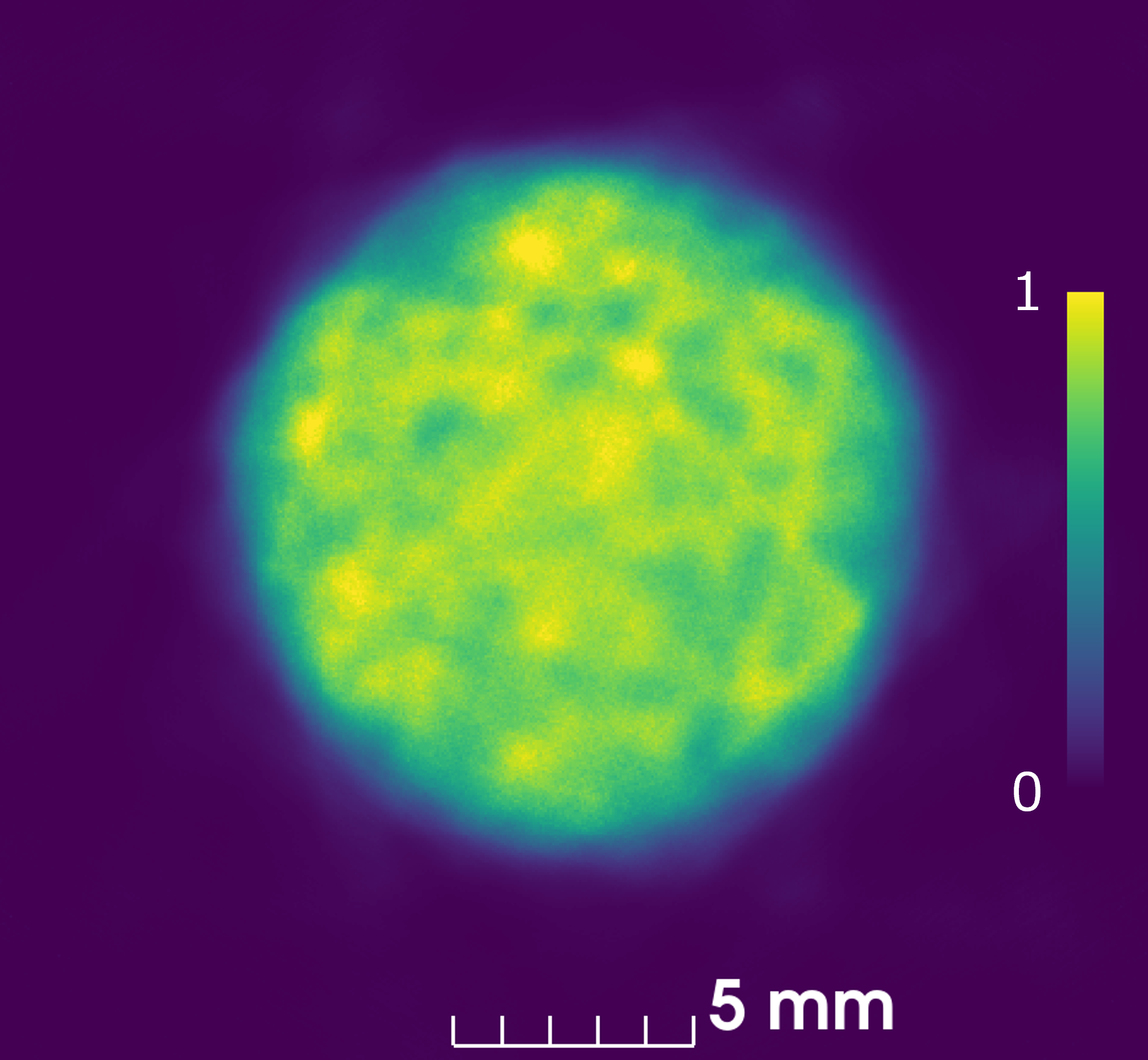}}
    \subcaptionbox{} {\includegraphics[width=0.49\linewidth]{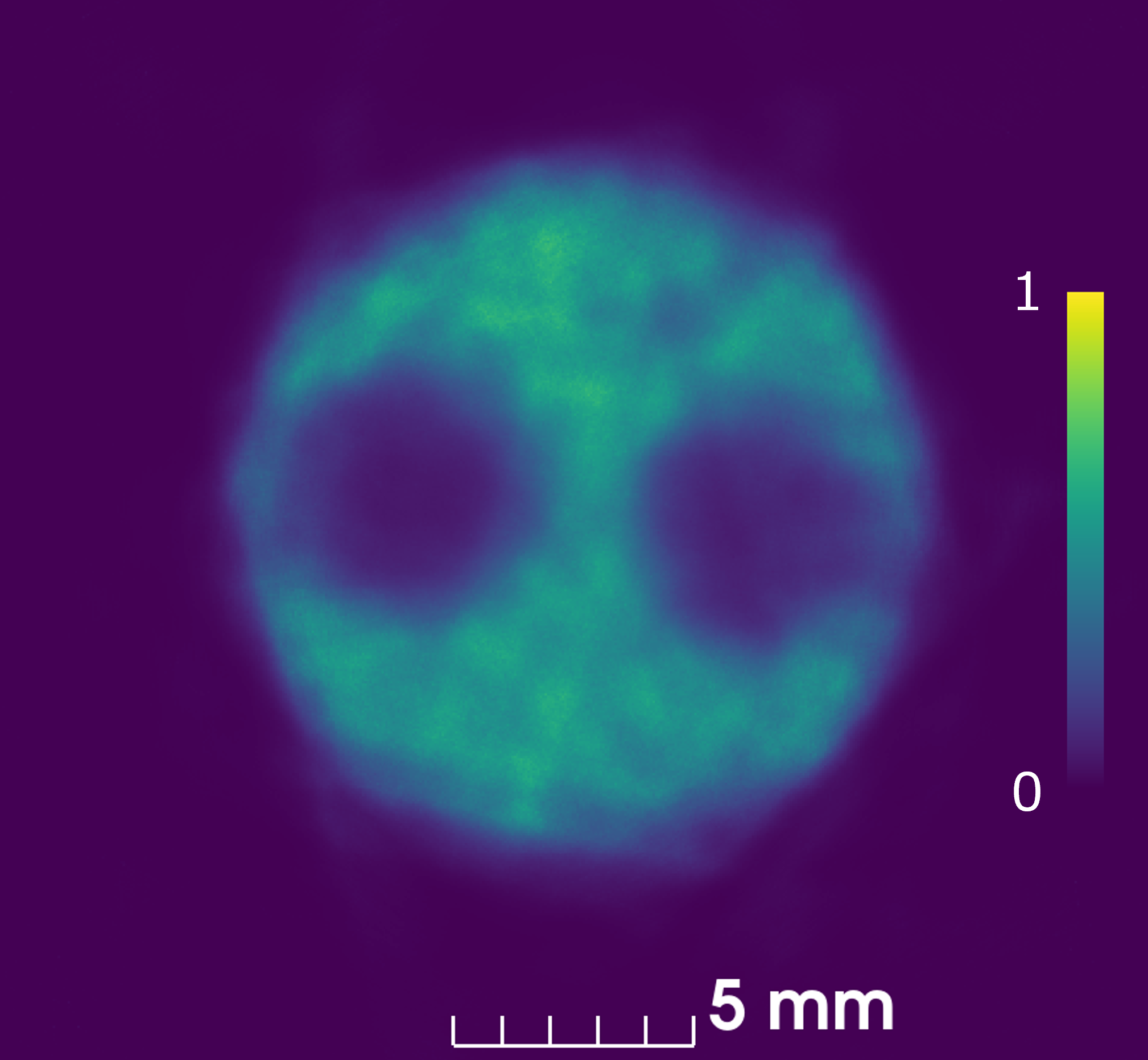}}
    \caption{(a-c) Projections over 5\,mm in each region of the reconstructed NEMA phantom from iteration 20. Four out of five rods are clearly visible.}
    \label{fig:NEMA_results_image}
\end{figure}

Fig.~\ref{fig:NEMA_results} (a) shows the RC values from the downsized NEMA phantom for iterations 5, 10, 15 and 20 and the four largest rods. Higher number of iterations (up to 50) did not lead to further detectable changes in the RC behavior, and therefore are not shown here. 
Uniformity and SOR are plotted in Fig.~\ref{fig:NEMA_results} (b-c). Projections of the reconstructed images for each section are presented in Fig.~\ref{fig:NEMA_results_image} (a-c).

\begin{figure}[htb]
    \centering
    \subcaptionbox{} {\includegraphics[width=0.25\textwidth]{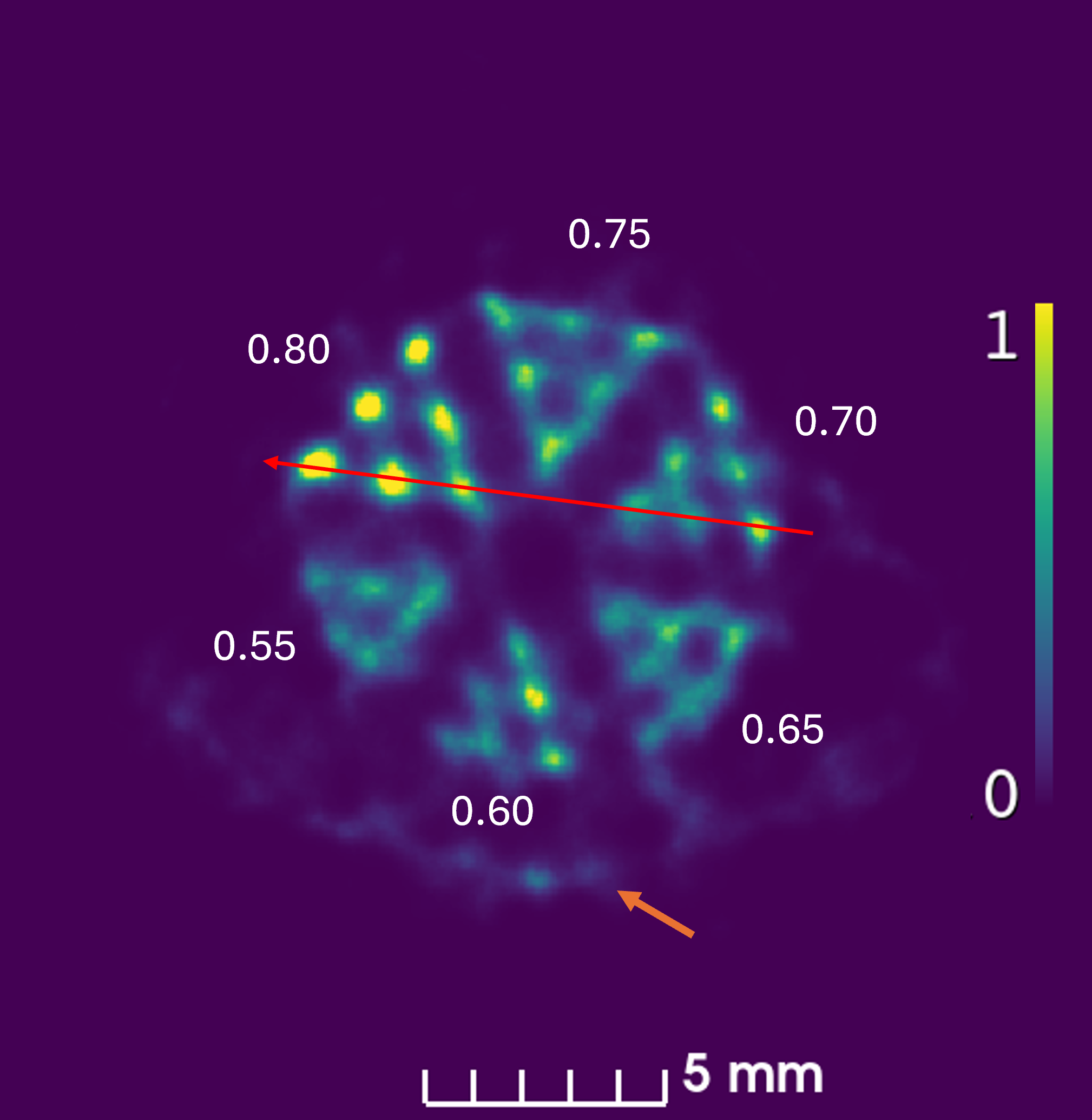}}
    \subcaptionbox{} {\includegraphics[width=0.35\textwidth]{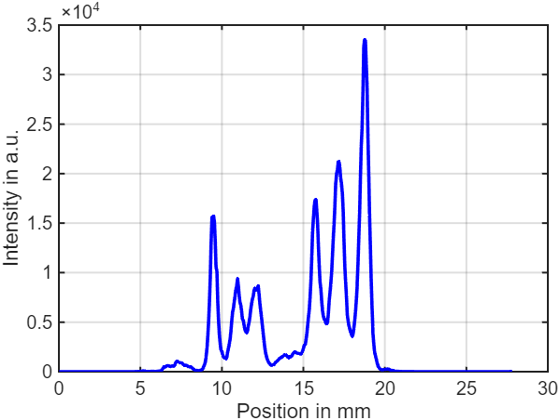}}
    \caption{(a) Summed slice of the reconstructed 3D-printed Derenzo phantom (projection over the central 6\,mm, iteration 100, filtered). The orange arrow indicates positron annihilations in the scanner bed. (b) Exemplary line profile (red line in (a)) between the 0.80 and the 0.70\,mm rod sectors of the projected central 6\,mm. }
    \label{fig:Derenzo_reco}
\end{figure}

Fig.~\ref{fig:Derenzo_reco}(a) shows the reconstructed image of the Derenzo phantom, while Fig.~\ref{fig:Derenzo_reco}(b) shows a selected line profile. Based on the VPR values computed from the projection over a segment of 6 mm along the rod lengths, 
(see Table \ref{tab:vpr_resolv}),
three out of six regions passed the 90\% criterion suggested by \cite{Hallen2020} as a measure of spatial resolution. 
This result indicates a spatial resolution of $0.7~\textrm{mm}$, which is consistent with visual inspection and the quantitative results obtained for the point-like source. Underneath the phantom, the scanner bed is visible; this is a result of the rods being only surrounded by air and the consequently extended positron range. The same effect with a similar phantom was also reported by \cite{Elmoujarkach_2024}.
\begin{table}[]
	\centering
	\caption[]{Mean VPR and resolvability of the sectors of the Derenzo phantom.}
	\begin{tabular}{lccc}
		\toprule
                        Rod size [mm]     &  VPR         & Resolvability \\ \midrule
0.80  & 0.18±0.06 & 100 $\%$    \\
0.75 & 0.35±0.07 & 100 $\%$  \\
0.70 & 0.24±0.05 & 100 $\%$   \\
0.65 &0.44±0.08 & 83.3 $\%$ \\
0.60 & 0.42±0.13& 83.3 $\%$ \\ 
0.55 & 0.75±0.11& 50 $\%$ \\ \bottomrule
	\end{tabular}
	\label{tab:vpr_resolv}
\end{table}

\subsection{Fish-like phantom}
\begin{figure}
    \centering
    \includegraphics[width=0.25\textwidth]{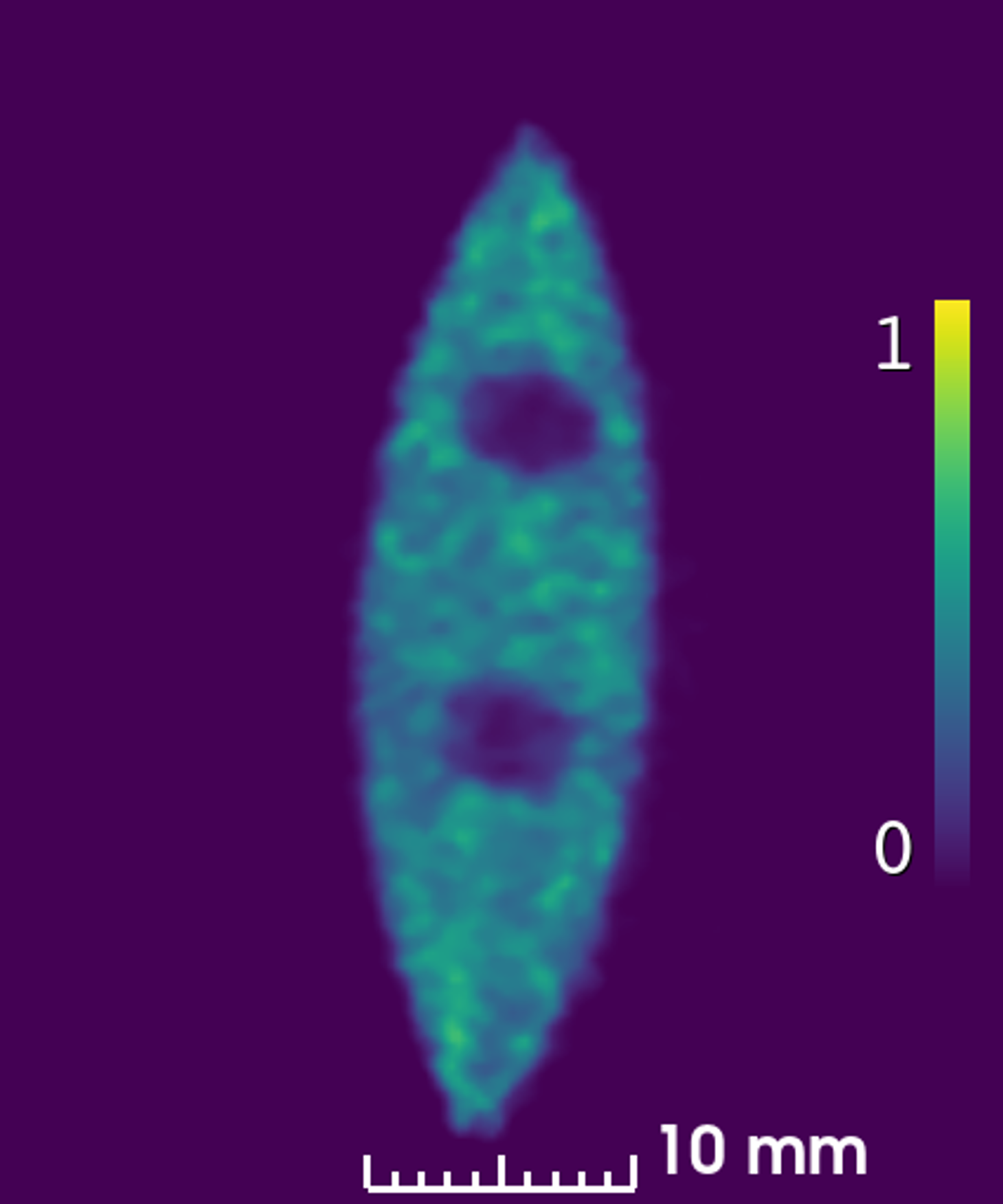}
    \caption{Coronal slice of the reconstructed 3D-printed fish-like phantom at iteration 20,
    filtered.}
    \label{fig:fishlike_reco}
\end{figure}

Fig.~\ref{fig:fishlike_reco} shows a central slice of the phantom reconstructed image.
Qualitatively, the reconstructed shape and the uniformity of the phantom body support the correct implementation of the decay correction and modeling of the axial bed shifts. 
\subsection{Ex-vivo \& in-vivo fish measurement}
\begin{figure}
    \centering
    \subcaptionbox{Coronal} {\includegraphics[width=0.45\linewidth]{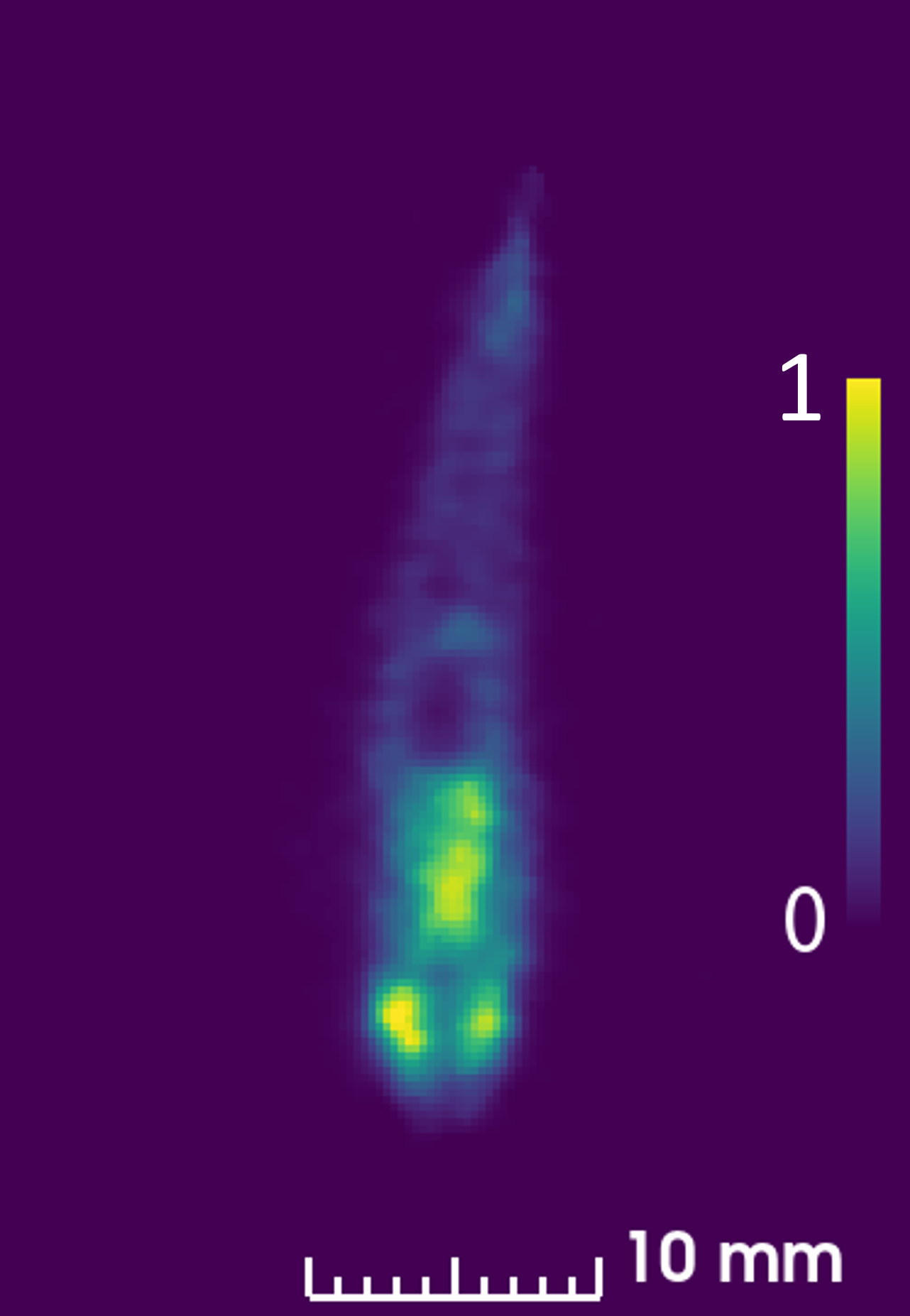}} 
    \subcaptionbox{Sagittal} {\includegraphics[width=0.45\linewidth]{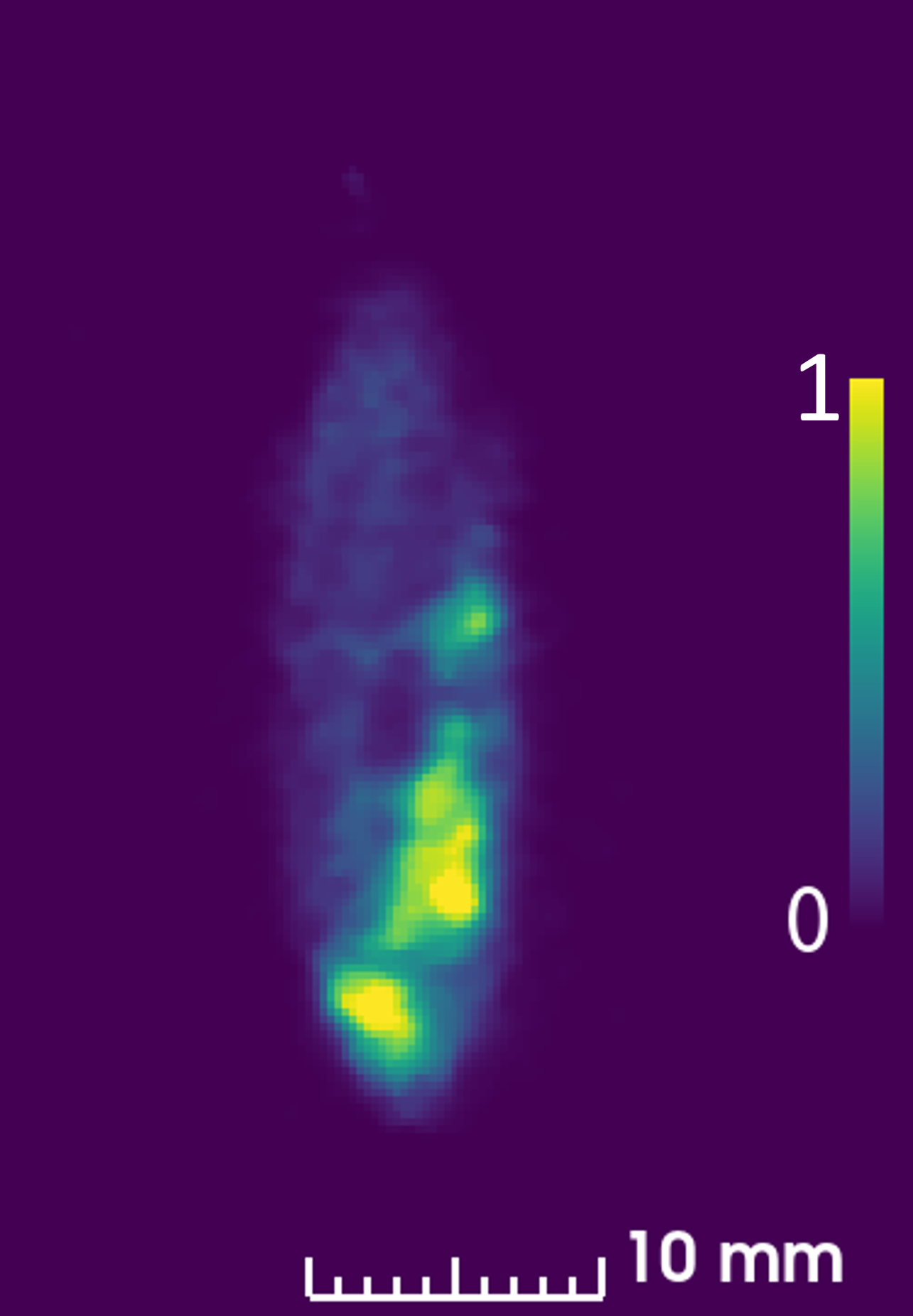}}
    \caption{Selected slices of the reconstructed image at iteration 20, filtered, of the ex-vivo adult zebrafish  scan with a start activity of 2.27\,MBq and a scan time of 62\,min.
    }
    \label{fig:exvivo_fish}
\end{figure}

\begin{figure}%[htb]
    \centering
    \subcaptionbox{Coronal} {\includegraphics[width=0.45\linewidth]{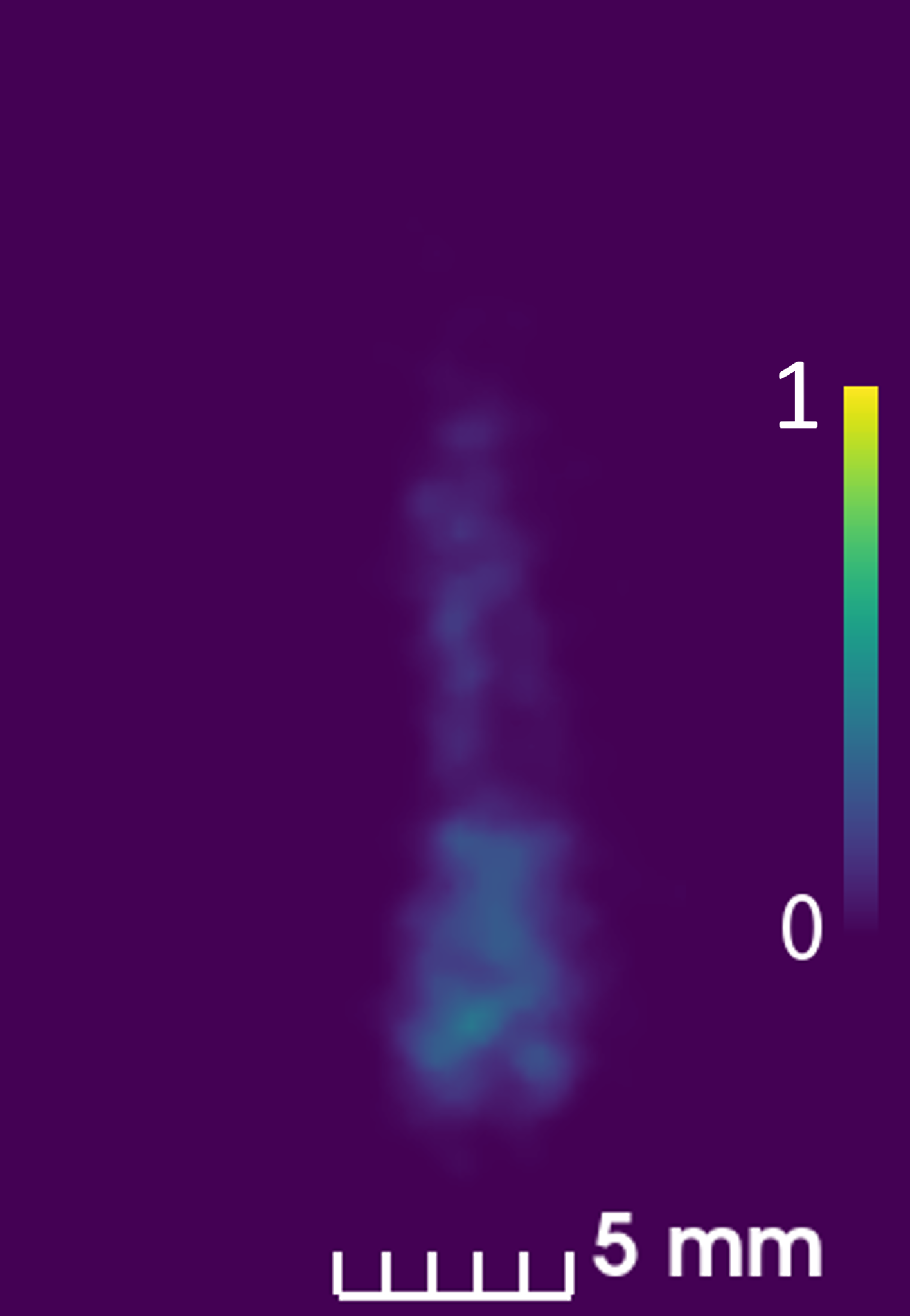}} 
    \subcaptionbox{Sagittal} {\includegraphics[width=0.45\linewidth]{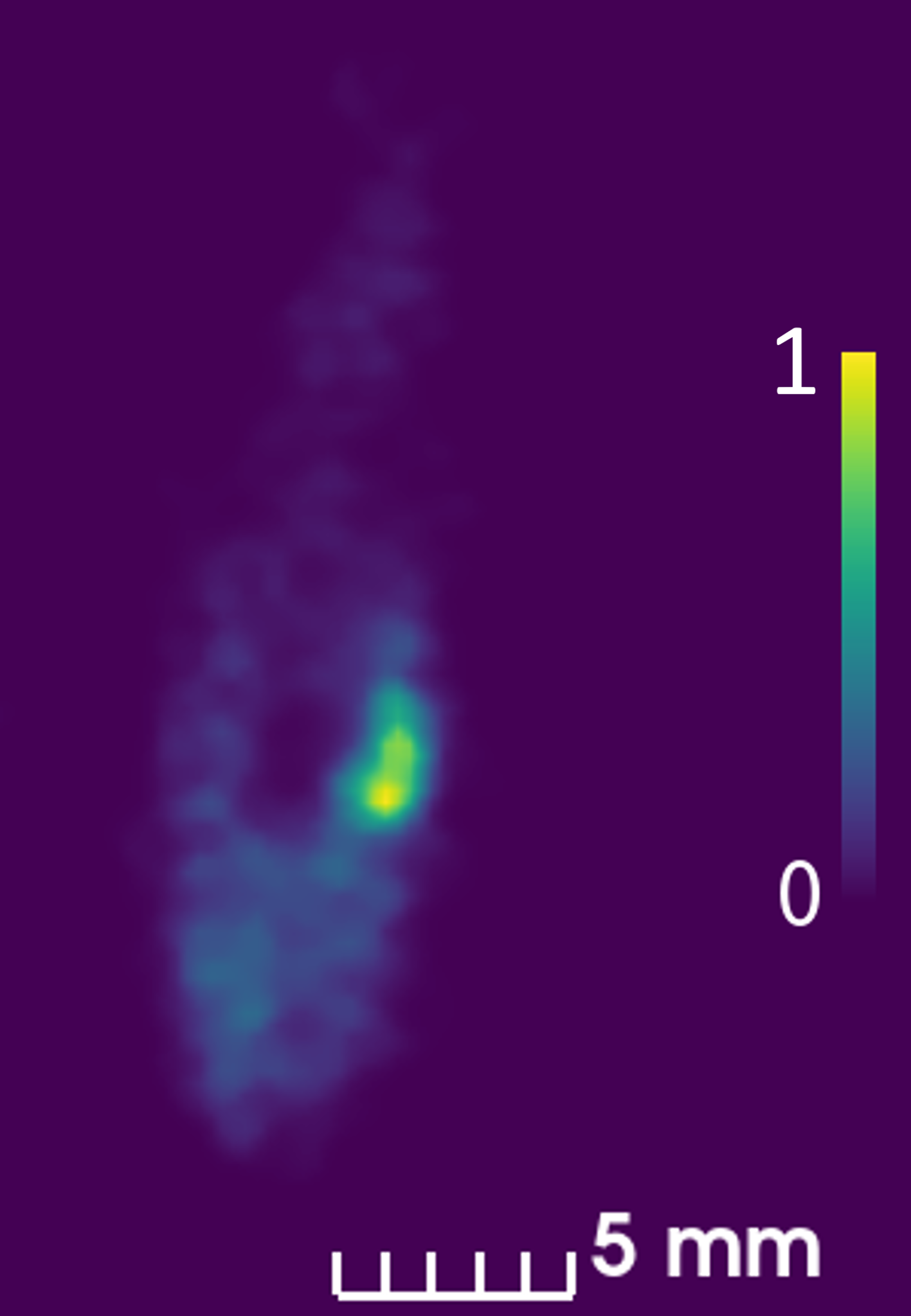}}
    \caption{Selected slices of the reconstructed image at iteration 20, filtered,  of the in-vivo adult zebrafish scan with a start activity of 0.62\,MBq and a scan time of 45\,min.}
    \label{fig:invivo_fish}
\end{figure}

Comparing the high activity ex-vivo scans (Fig.~\ref{fig:exvivo_fish}) with the low activity in-vivo scans (Fig.~\ref{fig:invivo_fish}), the effect of higher counts of the  ex-vivo measurement is evident. Note that the scaling is different, as the images are not intended for quantitative comparison.
In the coronal and sagittal image slices of the latter, various organs can be identified. 
In the coronal image, the eyes can be clearly distinguished from the inner organs, and the swim bladder is appreciable in the center (cold region). In the sagittal image, in addition to the injection site, the brain and a region of various organs (e.g., heart, liver) can be seen, although the latter cannot be separated from each other due to the nonspecific uptake of FDG in the fish. The cold swim bladder  is also clearly visible. The in-vivo scans mainly show  an uniform uptake in the fish body, the cold swim bladder and a higher uptake in location of the injection site.
Note that for both ex-vivo and in-vivo examinations, the fish were placed in water between tracer administration and the scan. 
The physical activity prior to the scan (swimming) was probably responsible for the uniform uptake in the trunk and posterior region, which comprises the musculature. 


\section{Discussion}
The results presented in this work validate the feasibility of the current prototype for small-fish imaging, while highlighting areas for future improvement.\\
The energy calibration measurements showed significant saturation effects at higher energies, which in turn is caused by the relatively low number of pixels per detector channel. In addition, a slight shift in the 511\,keV photopeak position was observed for all nuclides employed. This offset may be caused by inaccuracies (fit errors) when finding the peak positions of the three energies during the adjustment of the calibration function. However, it is also possible that the calibration formula used here (see Eq. \ref{eq:ekali}) cannot optimally reproduce the behavior of the detectors. As alternative, the calibration functions proposed in \cite{peng2019compton,nadig2025asics} could be tested in the future.

The mean energy resolution for \isotope[18]{F} was 21.6\% (see Table~\ref{tab:Eres}). The primary cause of this broadening is likely some optical crosstalk and the relatively low number of SiPM pixels per detector channel, which limits the dynamic range. For measurements involving the PET nuclide \isotope[89]{Zr}, MERMAID-v1 in its current form is not suitable due to the points just mentioned.
MERMAID-v1 energy resolution is worse than the one typically achieved by modern small-animal scanners, which often report values between 10\% and 15\% FWHM \cite{Krishnamoorthy2018,Wai2024}. 

The reconstructed images are not yet quantitatively accurate, as 
attenuation, scatter, efficiency and random corrections are still pending. 
Due to the small size of the objects and the amount of attenuating media, we believe that the impact of scatter events within the default energy window is not significant. Attenuation correction is also planned. As shown in \cite{Zvolsk2022}, even an attenuation map that only considers water in the imaging chamber should be sufficient to correct for attenuation. In any case, a CT system to provide an anatomical reference is under development, so attenuation factors will be obtained from the CT-based attenuation map. 

The count-rate performance measurements followed the expected exponential trend and indicated that MERMAID-v1 operates without significant dead-time effects or saturation for activities up to 18\,MBq. Although not relevant for the studied specimens with activities less than 3\,MBq, this might be essential for other in-vivo studies with higher radiotracer uptake. The underestimation of the calculated half-life extracted from the coincidence rates suggests the presence of a small number of random events. In future steps,  estimates of the random events, required for their correction in the reconstruction, will be extracted from the singles list-mode data as in \cite{oliver2016modelling}.

The absolute sensitivity at the FOV center of the system ranged from 0.06\% (default window) to 0.31\% (wide window). As expected, these values are lower than those of commercial ring-based small-animal PET scanners, which typically achieve sensitivities between 6 and 12\% \cite{Krishnamoorthy2018,Visser2009}, and are characterized by an axial extension of several centimeters. 
The low sensitivity of MERMAID-v1 is primarily due to its incomplete ring geometry and the short axial FOV. Concerning other research prototypes, in the work of Kang et al. \cite{Kang978}, a sensitivity of 1.56\% is achieved with an energy window of 440 to 560\,keV, despite having several complete detector rings. Similar values can be seen in the system by Wang et al. \cite{Wang2025}, reporting a sensitivity of 4.18\% for a large energy window of 350 to 650\,keV.
To increase the sensitivity, inter-crystal scatter events could be employed, as we have shown in \cite{roser2026}. In any case, the optimal energy window width should be determined in future studies with the upgraded system and with the scatter correction implemented.

The high spatial resolution is the most significant feature of the MERMAID-v1 prototype, in particular because it is achieved  without depth-of-interaction (DOI) information. In the central 12\,mm of the transaxial FOV, of particular relevance for zebrafish imaging, the average FWHM was $0.77 \pm 0.10~\textrm{mm}$ transaxially, and $0.66 \pm 0.10~\textrm{mm}$ axially (Table~\ref{tab:NEMA_FWHM_FWTM}).
Since these values were derived from an iterative reconstruction algorithm instead of filtered backprojection, the spatial resolution might be overestimated \cite{Gong_2016}. 
A degradation of the spatial resolution was observed towards the edges of the FOV, as expected, due to the scanner small diameter in combination with the lack of DOI information. The different behavior between the x-axis and y-axis shifts can be attributed to the lack of 90\textdegree\space symmetry, see Fig.\,\ref{fig:det_all}(c). Note that MERMAID-v1 incomplete ring geometry does not feature highly oblique crystal pair combinations, thus limiting the parallax effect. 

The visual inspection of Derenzo phantom projections allowed the 0.70 mm rods to be  identified, as well as some of the 0.6 mm rods. This sub-millimeter resolution is superior to commercial preclinical scanners, which typically offer resolutions around 1.0–1.5\,mm FWHM \cite{Visser2009, Rti2025}, and is comparable to specialized high-resolution research systems with DOI information. For example, the Molecubes beta-cube, which utilizes monolithic LYSO detectors, achieves a resolution of approximately 0.75\,mm FWHM at the center of the FOV \cite{Krishnamoorthy2018}. MERMAID-v1 attains comparable performance using a simpler rotating two-head design with pixelated crystals and no DOI information on the detector side, while still offering submillimeter resolution over the relevant FOV. This fact reflects the accurate mechanical alignment and the effective modeling of parallax effects included in the image reconstruction.
Higher resolution has been reached by the prototype developed by Kang et al. for mouse brain imaging, which achieved a resolution of 0.55 mm FWHM \cite{Kang978}; this value is competitive with autoradiography, as demonstrated in \cite{kang2026sub}.
The prototypes of Yang et al.  also show higher spatial resolution, namely 0.61 mm FWHM \cite{Yang2016} and 0.65 mm FWHM \cite{Yang_2026}. Additionally, both systems achieve higher sensitivity than the MERMAID prototype with sensitivities of 1.08 $\%$ and 14.9 $\%$ in the central FOV. 
Such a  level of performance, however, requires a sophisticated detector system based on a ring of staggered three-layer  crystal detectors or a more complex readout scheme.

Note that the purpose of MERMAID-v1 is not to compete with state-of-the-art high-resolution preclinical PET scanners, but to investigate the feasibility of a simplified PET system as a platform to establish protocols dedicated to zebrafish.

One  way of improving the spatial resolution would be to use a diffuse reflector for wrapping the scintillation crystals instead of a specular one. For instance, it has been shown that BaSO$_4$ can reduce optical crosstalk and improve crystal discrimination~\cite{Kang2021}, resulting in improved spatial resolution~\cite{Kang2025}; however, this may be accompanied by reduced light collection and degraded energy and timing performance~\cite{Kang2021}.

The quantitative assessment of image quality using the downscaled NEMA IQ phantom confirmed the system's ability to resolve small structures. 
However, the increase of the relative standard deviation in the uniform region with the iteration number (see Fig.~\ref{fig:NEMA_results} (b)) suggests a trade-off between contrast recovery and noise, which is typical for MLEM. In addition, a pronounced overshoot of the RC values was observed at higher iterations. This behavior can be attributed to the amplification of statistical noise in the iterative process of MLEM, in combination with low-count data. Furthermore, the NEMA protocol might be sub-optimal as it overestimates systematically the RCs, especially in high-variance cases, see \cite{Hallen2020}. 
Noise control through regularization can be a further option to improve imaging performance at the software level, a strategy also adopted by commercial small-animal systems (e.g., \cite{nagy2013performance}).
The SOR for air-filled cavities was low (0.09 at iteration 20), indicating a low scatter contribution, likely due to the small size of the phantom and the absence of water in the phantom setup (in contrast to the in-vivo case). For this phantom, a comparison to other existing systems is not meaningful due to its reduced size. 

The successful imaging of adult zebrafish marks a critical milestone for the MERMAID project. The ex-vivo images (Fig.~\ref{fig:exvivo_fish}) clearly displayed a higher tracer uptake in the eyes, brain, and internal organs, as well as no uptake in the swim bladder. While the in-vivo scan (Fig.~\ref{fig:invivo_fish}) was limited by the low tracer uptake, it still demonstrated the potential of the system to provide useful insights of an anesthetized fish placed in a water-filled chamber. These initial results indicate the need to improve tracer application methods (e.g., microinjectors) and to refine imaging protocols. Particularly, and as mentioned above, the lack of data corrections prevents absolute quantification. Despite this limitation, the current prototype could allow for qualitative lesion detection tasks, and qualitative longitudinal scans. Dynamic studies would not only require quantitative images, but also improvements in the design, optimization of tracer delivery and refinement of imaging protocols.

\section{Conclusion \& outlook}
This study shows the feasibility of dedicated PET imaging for adult zebrafish and similar small aquatic animals using MERMAID-v1.
With submillimeter spatial resolution, acceptable energy resolution, and the ability to perform in-vivo scans under physiological conditions, MERMAID-v1
represents a significant step toward addressing existing limitations in zebrafish imaging.
While sensitivity remains lower than full-ring PET systems, the current design 
offers advantages in animal welfare and experimental flexibility.
However, the present setup is not final and will be refined in future versions.
Next steps will focus on implementing efficiency, scatter, random and attenuation corrections, and expanding the system to a total of eight detector modules of the same type. This will lead to higher sensitivity and shorter measurement times while keeping the flexibility of the system. To allow for such an extension of the prototype, a modification of the PCB design will be necessary.  
The integration of a low dose CT modality is envisioned. It will not only add anatomical guidance but will also be used for attenuation correction.
Ultimately, MERMAID-v1 contributes to extend PET by enabling in-vivo imaging of adult zebrafish. Beyond its biomedical applications, such a dedicated PET platform could also provide useful insights into the physiology and disease of other small aquatic animal species, making it a valuable tool for marine biology, and environmental research.

\section{Acknowledgments}
This work was partly supported by the German Research Foundation (DFG) under grant agreement no. 496099829 and the DFG Cluster of Excellence PMI under grant agreement no. 390884018. The authors gratefully acknowledge the computing time made available to them on the high-performance computer ‘Lise’ at the NHR center NHR@ZIB, Project No. shb00004. This center is jointly supported by the Federal Ministry of Education and Research and the state governments participating in the NHR (www.nhr-verein.de). J. Roser was partly supported by the Alexander von Humboldt Foundation and currently by the European Union’s Horizon 2024 research and innovation programme under the Marie Skłodowska-Curie grant agreement no. 101207067.

The authors would like to thank the Department of Radiology and Nuclear Medicine of the Universitätsklinikum Schleswig Holstein (UKSH) Campus Lübeck for providing the radiotracer, also Luise Morgner, Siobhan McIntyre, and Alida Nestler who helped  testing and preparation of the 3D-printed phantoms, Cindy Läken for experimental help and Dirk Steinhagen for technical support throughout the entire MERMAID project.

\bibliographystyle{IEEEtran}
\bibliography{ref}

\end{document}